\documentclass{aa}

\usepackage{graphicx}
\usepackage{natbib}
\usepackage{siunitx}
\usepackage[utf8]{inputenc}
\usepackage{threeparttable}
\usepackage{placeins}
\usepackage{hyperref}
\usepackage{tablefootnote}
\usepackage[normalem]{ulem}

\DeclareRobustCommand{\rchi}{{\mathpalette\irchi\relax}}
\newcommand{\irchi}[2]{\raisebox{\depth}{$#1\chi$}} 

\usepackage[varg]{txfonts}
\def \bp{$\beta$\,Pic}
\def \betapictoris{$\beta$\,Pictoris}

\def\feii{Fe\,{\sc ii}}
\def\fei{Fe\,{\sc i}}
\def\mnii{Mn\,{\sc ii}}

\def\caii{Ca\,{\sc ii}}
\def\aliii{Al\,{\sc iii}}
\def\siiv{Si\,{\sc iv}}
\def \A{$\si{\angstrom}$}

\begin{document}

\title{Curves of growth for transiting exocomets: \\ Application to Fe\,{\sc ii} lines in the $\beta$\,Pictoris system} 
\titlerunning{Transiting exocomets curve of growth}

\author{
T.\ Vrignaud\inst{1}
\and
A. Lecavelier des Etangs\inst{1}
\and
F. Kiefer\inst{2} 
\and 
A.-M.~Lagrange\inst{2}
\and
G.~H\'ebrard\inst{1}
\and
P. A.~Str\o m \inst{3}
\and
A.~Vidal-Madjar\inst{1}
}

\institute{
Institut d'Astrophysique de Paris, CNRS, UMR 7095, Sorbonne Université, 98$^{\rm bis}$ boulevard Arago, 75014 Paris, France 
\and 
LESIA, Observatoire de Paris, Université PSL, CNRS, 5 Place Jules Janssen, 92190 Meudon, France
\and
Department of Physics, University of Warwick, Coventry CV4 7AL, UK
}


\abstract{ 
Using transit spectroscopy, exocomets are routinely observed in the young planetary system of  \bp. However, despite more than 35~years of observations, we still have very little information on the physical properties and almost no information on the abundances of the gaseous clouds surrounding the comets' nuclei, the difficulty being the conversion of the observed absorption profiles into column density measurements.

Here, we present a new method to interpret the exocomet absorptions observed in \bp\ spectrum and link them to the physical properties of the transiting cometary tails (e.g. size, temperature, and column density). We show that the absorption depth of a comet in a set of lines arising from similar excitation levels of a given chemical species follows a simple curve as a function of $g$$\cdot$$f$, where $f$ is the line oscillator strength and $g$ its lower level multiplicity. This curve is the analogue of the curve of growth for interstellar absorption lines, where equivalent widths are replaced by absorption depths. To fit this exocomet curve of growth, we introduced a model where the cometary absorption is produced by a homogeneous cloud, covering only a limited fraction of the stellar disc. This model is defined by two parameters: $\alpha$, characterising the size of the cloud relative to the star, and $\beta$, related to the optical depth of the absorbing gas.

This model was tested on two comets observed with the Hubble Space Telescope in December 1997 and October 2018, in a set of lines of ionised iron (\feii) at 2750 \A. The measured absorption depths are found to satisfactory match the two-parameter curve of growth model, indicating that both comets cover roughly 40 \% of the stellar disc ($\alpha = 0.4$) and have optical thicknesses close to unity in those lines ($\beta \sim 1$).

Then, we show that if we consider a set of lines arising from a wider range of energy levels, the absorbing species seems to be populated at thermodynamical equilibrium, causing the cometary absorption to follow a curve of growth as a function of  $g f \cdot e^{-E_l/k_B T}$ (where $T$ is the temperature of the absorbing medium). For the comet observed on December 6, 1997, we derive a temperature of $T=10\,500\pm 500$\,K  and a total \feii\ column density of $N_{\rm FeII}=(1.11 \pm 0.09)\times 10^{15}$cm$^{-2}$. By considering the departure from the Boltzmann distribution of the highest excited energy levels ($E_l\sim 25\,000$\,cm$^{-1}$), we also estimate an electronic density of $n_e\approx (3 \pm 1) \times 10^{7}$\,cm$^{-3}$.
}


\maketitle

\section{Introduction}
The star \bp\ is a young nearby star of about 20 million years old \citep{Miret-Roig_2020}, located 19.3\,parsecs away from us and harbouring a gaseous and dusty debris disc seen edge-on \citep{Smith_1984, Vidal-Majar_1986, Kalas_2001, Roberge_2006,Apai_2015,Brandeker_2016}. Two massive planets were discovered through imagery, radial velocity (RV), and mutual dynamical perturbations \citep{Lagrange_2010,Snellen_2018, Lagrange_2019, Nowak2020, Lacour_2021}. Therefore, \betapictoris\  offers a unique opportunity to investigate the early stages of a planetary system. 

For more than thirty years, exocomets transiting the star have been detected using spectroscopy, probing the gaseous part of the cometary comas and tails \citep{Ferlet_1987,Beust_1990,Vidal-Madjar_1994,Kiefer_2014, Kennedy_2018}. 
First named by the circumlocution falling evaporating bodies (FEB, see \cite{Beust_1990,Vidal-Madjar_1994}), exocomets in \bp\ are the analogous of comets in our own Solar System, that is, icy bodies whose volatiles evaporate when approaching the star \citep{Strom_2020}, producing huge gaseous and dusty tails that can be detected using transit spectroscopy and transit photometry \citep{Zieba_2019, Pavlenko_2022, Lecavelier_2022}. One of the most remarkable aspects of the exocomet population in the \bp\ system is its unique accessibility to detailed analysis. The close proximity of this young star, coupled with orientation favouring transits of bodies with low inclination relative to the disc, allows us to scrutinise this exocomet population with a level of detail currently unattainable for most other systems. Although some systems seem to offer interesting possibilities for similar studies, such as 49~Cet \citep{Roberge_2014, Miles_2016},  HD\,172555 \citep{Kiefer_2014b, Kiefer_2023}, and a few others identified by surveys \citep{Montgomery2012,Welsh2013,Rebollido2020}, the number of comets in the \bp\ system is much larger than in any other known exocometary system. \bp\ also remains the only system for which the presence of exocomets has been linked to the presence of cold CO that, in the absence of $H_2$, must be continuously resupplied \citep{Vidal-Madjar_1994,Jolly1998,Roberge2000, Lecavelier2001}, most likely by evaporating comets.

\begin{table*}[h!]
        \centering      
        \caption{\centering Log of HST observations used in the present study.}

        \small \renewcommand{\arraystretch}{1.4} \begin{tabular}{| c | c | c | c | c | c |}          
                \hline                     
                  Spectrum & Programme ID & Date & Wavelength (\A) & Exposure time (s) & Grating\\    
                \hline  
o4g001060  & 7512 & 1997-12-06 & 2128-2396 & 288 & E230H   \\
                \hline
o4g001050  & 7512 & 1997-12-06 & 2377-2650 & 360 & E230H   \\
                \hline
o4g001040  & 7512 &  1997-12-06 & 2620-2888 & 360 & E230H   \\
                \hline
o4g002060  & 7512 &  1997-12-19 & 2128-2396 & 288 & E230H   \\
                \hline
o4g002050  & 7512 &  1997-12-19 & 2377-2650 & 360 & E230H    \\
                \hline
o4g002040  & 7512 &  1997-12-19 & 2620-2888 & 360 & E230H   \\
                \hline

odaa01010  & 14735 & 2017-10-09 & 2665-2931 & 475 & E230H   \\
                \hline
odaa03010  & 14735 & 2018-01-10 & 2665-2931 & 475 & E230H   \\
                \hline
odaa62010  & 14735 & 2018-03-07 & 2665-2931 & 475 & E230H   \\
                \hline

odt901010  & 15479 & 2018-10-29 & 2665-2931 & 2101 & E230H    \\
                \hline
odt902010  & 15479 & 2018-12-15 & 2665-2931 & 2101 & E230H   \\
                \hline 

        \end{tabular}   
    \label{recap_observations}
\end{table*}

However, despite the numerous observations of exocometary transits in the \bp\ system, most previous studies focussed on modelling the observed spectroscopic events to conclude on the exocometary nature of the phenomenon \citep[][]{Lagrange_1989, Vidal-Madjar_1994},  numerical simulations \citep[e.g.][]{Beust_1990, Beust_1993}, the detection of the widest variety of species possible \citep[e.g.][]{Vidal-Madjar_1994}, or the derivation of their orbital properties \citep[e.g.][]{Kiefer_2014, Kennedy_2018}. As a result, only very few studies concluded on the measurement of physical properties of the observed comets themselves, with notably the work by 
\cite{Lagrange_1995} for first estimates of column densities using equivalent widths,  \cite{Mouillet_1995} for temperature and density of the gaseous coma, \cite{Kiefer_2014} for the estimate of the nuclei evaporation efficiency, and the analysis of the photometric transits by \cite{Lecavelier_2022} for the size distribution of the cometary nucleus.  This situation results from the extreme difficulty in converting the observed absorptions into intrinsic physical quantities of the transiting comets. Here, using the advantage of a large number of electronic transition lines of the same species in a few exocomets observed with the Hubble space telescope, we aim to provide a detailed analysis of the transit spectra in order to better characterise the physical conditions in the gaseous clouds surrounding the comets' nuclei.

The selection of the observational data and their analysis are described in Sect.~\ref{Observations}. The first model and the derivation of the exocomet curves of growth are discussed in Sect.~\ref{First Model}. The estimates of the column densities, temperature, and electron density are presented in Sect.~\ref{validation 2 parameters} and Sect.~\ref{Temperature and electron density}. We discuss, summarise, and conclude in Sect.~\ref{Discussion} and Sect.~\ref{Conclusion}.

\section{Observations}
\label{Observations}

\subsection{Archival data}

For our study, we used mostly unpublished archival UV spectra of \betapictoris, obtained with the Space Telescope Imaging Spectrograph (STIS) onboard the Hubble Space Telescope (HST) within three different guest-observer programmes (programme \#7512, PI A.-M.~Lagrange, \#14735 and \#15479, PI F.~Kiefer; see Table~\ref{recap_observations}). For programmes \#14735 and \#15479 (executed in 2017/2018), no analysis of the data has yet been published. The data of programme \#7512 (December 1997) was analysed in \cite{Roberge2000} to constrain the CI and CO content of the circumstellar disc, but no study of the exocometary absorption was performed.

These STIS observations were obtained using the high-resolution ($R\sim 114\,000$) echelle grating {\it E230H} at seven~different epochs. The raw data were processed using version 3.4.2 of the {\tt calstis} pipeline released on January 19, 2018, providing 1-D flux calibrated spectra that are publicly available and have been downloaded from the MAST archive. The data are wavelength calibrated on a heliocentric reference frame.

All these spectra cover the ionised iron (\feii) line series between 2715 and 2769 \A, rising from the 3d$^6$($^5$D)4s - a$^4$D metastable levels, in which cometary transits are easily detected (see sect. 2.3). In addition, the two 1997 observations (programme \# 7512) also cover the \feii\ line series at 2400 \A\ and 2600 \A, which rise from a wide range of energy levels.

\begin{figure*}[hbtp]
    \centering    
    \includegraphics[clip, scale = 0.172, trim = 400 50 520 20]{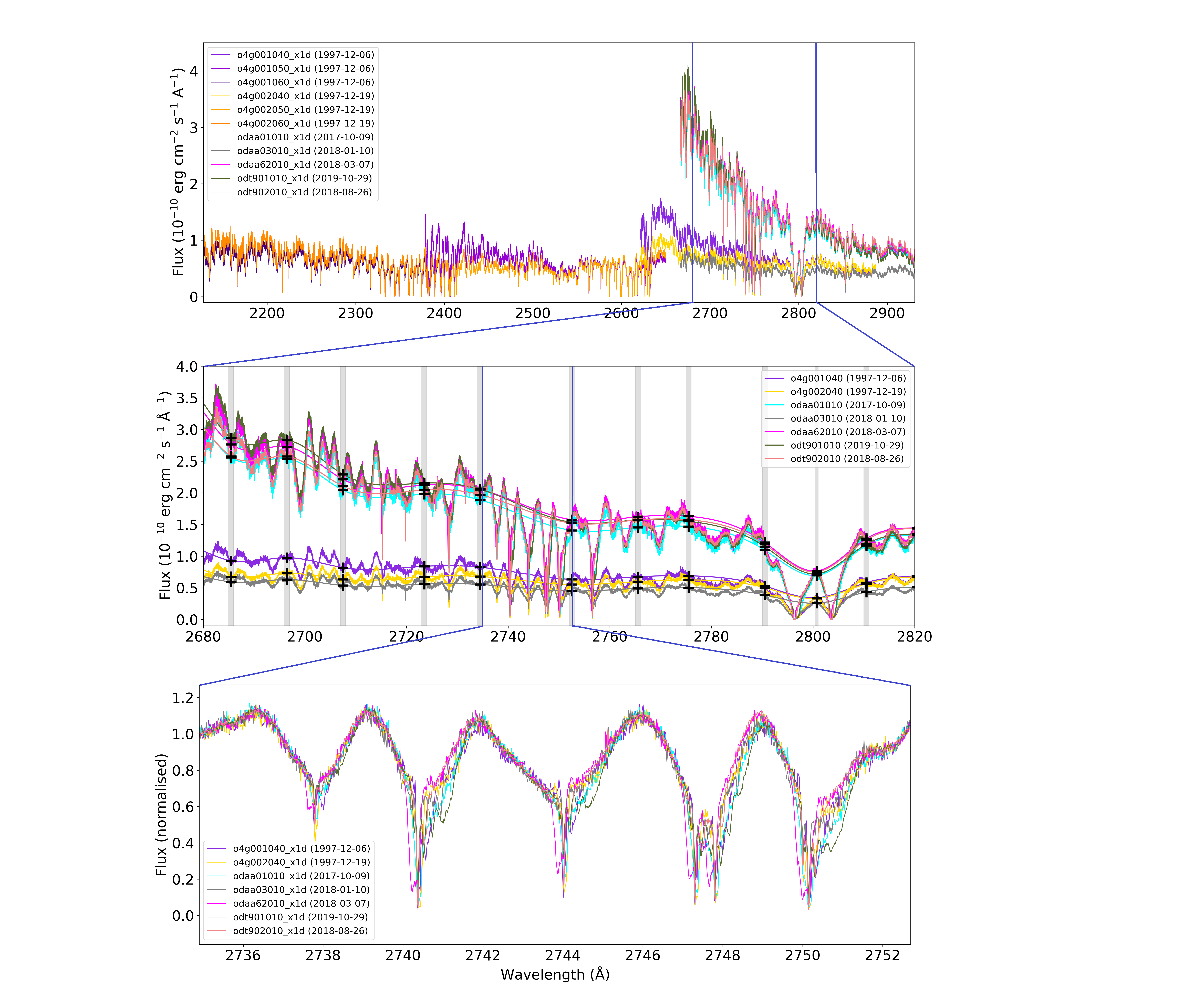}

    \caption{Illustration of the recalibration method. 
    \textbf{Top panel}: Raw spectra show significant shape disparities, probably resulting from changes in instrument calibration throughout the years. These calibration disparities are hardly fitted by polynomials. 
    \textbf{Centre panel}: Zoomed-in view of 2680-2820 \AA\ region, showing the splines fitted to each spectrum. The grey areas indicate the different spectral domains where the reference fluxes were measured (black crosses), which are then used to calculate a cubic spline for each spectrum. 
    \textbf{Bottom panel}: Zoomed-in view of 2735-2753 \A\ region after renormalisation by the cubic splines. The spectra taken at different epochs are shown with different colours; they are well superimposed in the region of the stellar continuum, while variable absorptions are clearly visible around the different \feii\ lines between 2737 and 2750 \A .}
    \label{flux_renorm}
\end{figure*}

\subsection{Data analysis}

Each of the STIS/E230H spectra covers a bandwidth of about 270\,\AA\ wide, with a set of 24 to 40 orders of about 12\,\AA\ wide, partially overlapping. For each exposure, the fluxes of all the orders have been combined into a single spectrum and resampled on a common wavelength table with a resolution of 18 m\AA. The wavelengths were also corrected for the heliocentric radial velocity of \bp, +20 km/s \citep{Gontcharov2007}, so that they correspond to the \bp\ reference frame.

In a second step, to correct for variation of the flux calibration from one epoch to another, we renormalised the measured flux to a common flux level, independent of exocomet activity. For that, we selected a set of 65 $\sim$1\A-wide regions free of any variations (i.e. mostly outside spectral lines) separated by 10 to 20\,\AA. By measuring their flux on these different regions, we obtained, for each spectrum, a set of reference points of the form $(\lambda_i, f_i)_{i=1,2...}$. The reference points of each spectrum were then fitted by cubic splines, and each spectrum was divided by its corresponding spline. This flux normalisation allowed us to obtain data that can be easily compared, unveiling variable absorption features around many lines (Fig.~\ref{flux_renorm}).

Finally, as no significant spectral variation was observed within each set of three exposures taken on December~6 and December~19, 1997 (both times within a single HST orbit of $\sim 90$ minutes), these sets of exposures were merged together, yielding a single spectrum from 2128\,\AA\ to 2888\,\AA\ for each of the two epochs. For the 2017-2018 observations, no such merging was necessary, as only one exposure was used for each visit. As a result, we ended up with seven normalised spectra: two spectra covering the 2128\,\AA\ to 2888\,\AA\ wavelength range taken 13~days apart in 1997, and five spectra covering the 2665\,\AA\ to 2931\,\AA\ wavelength range taken at five different epochs in 2017-2018.

\subsection{Detected \feii\ lines}

Dozens of \feii\ lines are detected in the seven analysed spectra, with deep absorption originating both from the circumstellar disc (stable) and from transiting evaporating comets (variable).  
These exocomet signatures are detected in three \feii\ main series: a first one near 2750 \A, where the high number of different visits (7) allows for distinguishing the variable absorptions from the stable stellar spectrum fairly easily (Fig. \ref{Fe II lines 2750}), as well as two other groups near 2400 and 2600 \A, only covered by the two 1997 observations (Fig.~\ref{Fe II lines 2600}). In addition, a few lines arising from highly excited levels also showed fainter variable absorptions, as seen in Fig.~\ref{Fe II lines faint}. The complete list of lines used in the following sections is presented in Tables~\ref{list lines} (for the strongest lines) and~\ref{list lines faint} (for the lines arising from high energy levels).

\subsection{Reconstruction of the comet-free spectrum}
\label{reconstruction of the reference}

\begin{figure}[h]
 \centering
 \includegraphics[clip, trim = 48 40 50 70, scale = 0.36]{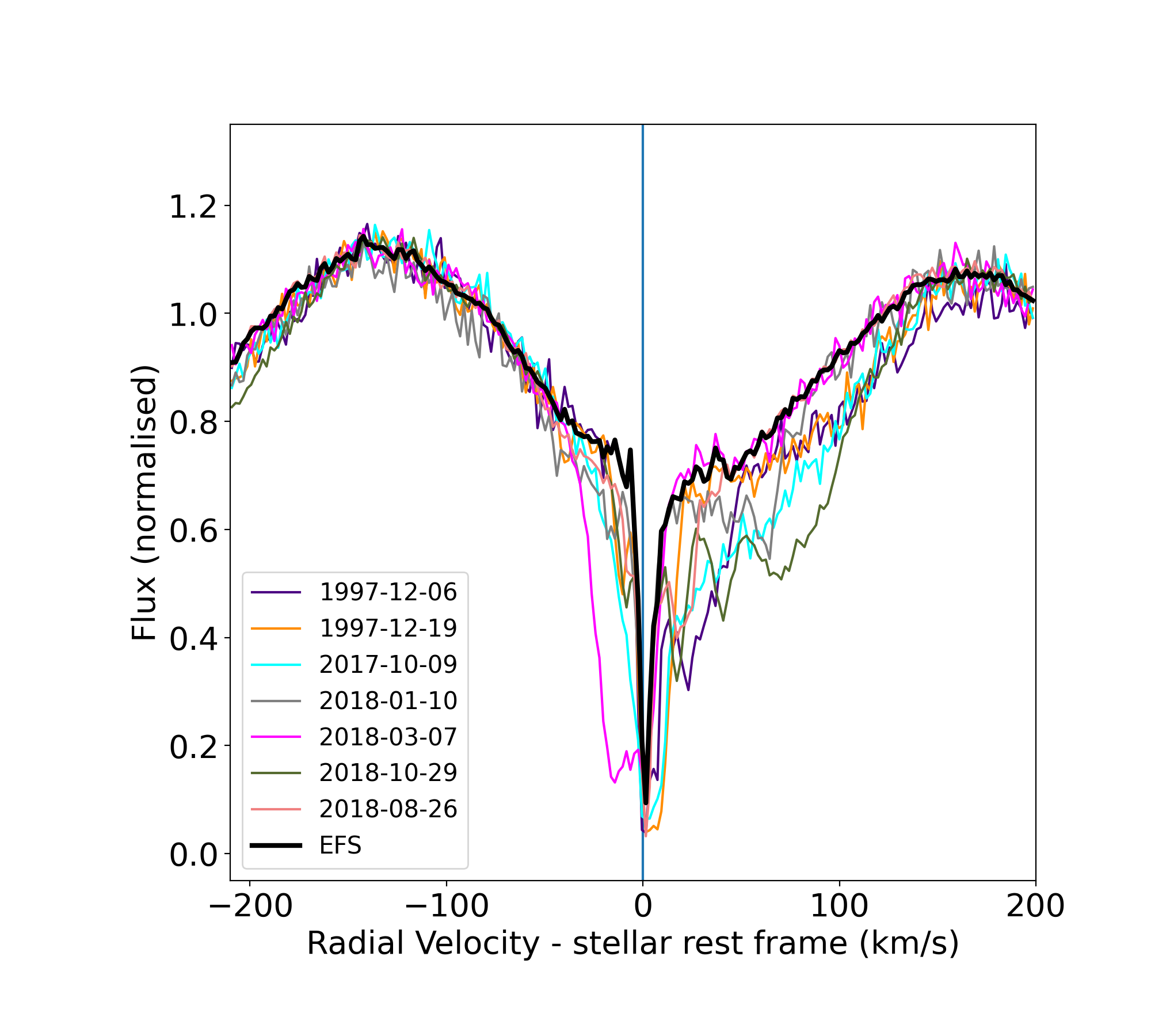}
      \caption[]{Variable absorptions in the 2740 \A\ \feii\ line. 
      The black line shows the reference comet-free spectrum of \bp, obtained by averaging the unabsorbed spectra at each individual wavelength. The deep and narrow absorption near 0 km/s is due to the stable component of the circumstellar disc of \bp,\, which could be slightly blueshifted compared to the stellar radial velocity (see \cite{Brandeker2011}).
    }
 \label{2740_master}
\end{figure}

In order to quantify the variable features observed in these many \feii\ lines, particularly in the series at 2750 \A, one must first compute a stellar reference spectrum without cometary absorption. To do so, we proceeded in two main steps:

\begin{itemize}
    \item First, we used the $\sim$2750\,\A\ \feii\ lines to identify, for each spectrum, the radial velocity domains where cometary absorption occurs, assuming it always occurs at the same radial velocities from one line to another (this assumption is in fact easily verified; see Sect. \ref{Cometary absorption}). This identification is made possible by the large number of spectra at these wavelengths, which allows us to easily distinguish, for each observation, the spectral domains where it is affected by cometary absorption and the spectral domains where it is not (see, e.g. Fig.~\ref{Fe II lines 2750}). More quantitatively, for each radial velocity $v$, we considered that a given spectrum is affected by cometary absorption when its flux at the velocity $v$ in the 2740 and 2756~\A\ lines (those are the strongest lines of the 2750~\A\ series) is at least 5 \% below the brightest observation. The list of the radial velocity domains where cometary absorptions are detected is given for each observation in Table~\ref{RV_ranges_comet}: we note that, for each radial velocity, there is always at least one comet-free observation, thanks to the rather high number of independent visits.

    \begin{table}[h!]
        \centering  
        
        \small \renewcommand{\arraystretch}{1.3} 
        
        \begin{threeparttable}

            \caption{List of radial velocity ranges where cometary absorption is detected, for each observation.}
            
        \begin{tabular}{| c | c |}          
                
            \hline                     
            Date of observation & RV ranges with cometary absorption (km/s) \tnote{a}\\        
            \hline  
            
            1997-12-06 & [-4; +200]\\
            1997-12-19 & [-17; +19 ], [+50; +150] \\
            2017-10-09 & [-35; +120]\\
            2018-01-10 & [-70; 0], [+19; +150] \\
            2018-03-07 & [-40; +13]\\
            2018-10-29 & [-30; -4], [+8; +140 ] \\
            2018-12-15 & [-40; +50]\\
            \hline
        \end{tabular}   

        \begin{tablenotes}
            \item[a] The radial velocities are given in \bp\ rest frame.
        \end{tablenotes}
    \end{threeparttable}

    \label{RV_ranges_comet}
\end{table}

    \item Second, for each wavelength pixel, we used the list of the radial velocity domains where comets are detected in each spectrum (Table~\ref{RV_ranges_comet}),  and the list of all the \feii\ lines (Tables~\ref{list lines} and \ref{list lines faint}), to determine the set of observations that can be considered comet-free at this particular wavelength. We then averaged all these un-absorbed spectra (free of cometary absorption) to obtain the reference spectrum of \bp. For instance, for the line at 2740\,\AA\ (Fig. \ref{2740_master}), the comet-free spectrum between +50 and +120 km/s is obtained by averaging the 2018-03-07 and 2018-12-15 observations.
\end{itemize}

We note in Fig.~\ref{2740_master} that it is not completely clear weather the spectrum obtained through this calculation is indeed the \bp\ comet-free spectrum. For instance, shallow comets may have been missed in our identification, leading us to underestimate the reference spectrum by a small fraction (typically a few percent). On the other hand, the reference spectrum might be slightly overestimated near + 30 km/s because of higher flux at the corresponding wavelengths in the March 7, 2018 observation (see Fig. \ref{2740_master}, pink line). However, the impact of these biases on the calculation of the comet-free spectrum is much smaller than the typical signature of a \bp\ exocomet on its host star spectrum (which can cause absorption depths as high as 50 \%). The conclusions presented in this paper are thus not significantly affected by such systematic errors.

It should also be noted that this comet-free spectrum can still contain interstellar and circumstellar absorption. However, because these absorptions are stable, they do not impact the measurement of cometary absorption depth. They can thus be included in the stellar reference spectrum, as if they had a stellar origin.

\subsection{Cometary absorption}
\label{Cometary absorption}

\begin{figure}[h!]
 \centering
  \includegraphics[clip, trim = 0 0 0 50 , scale = 0.4]{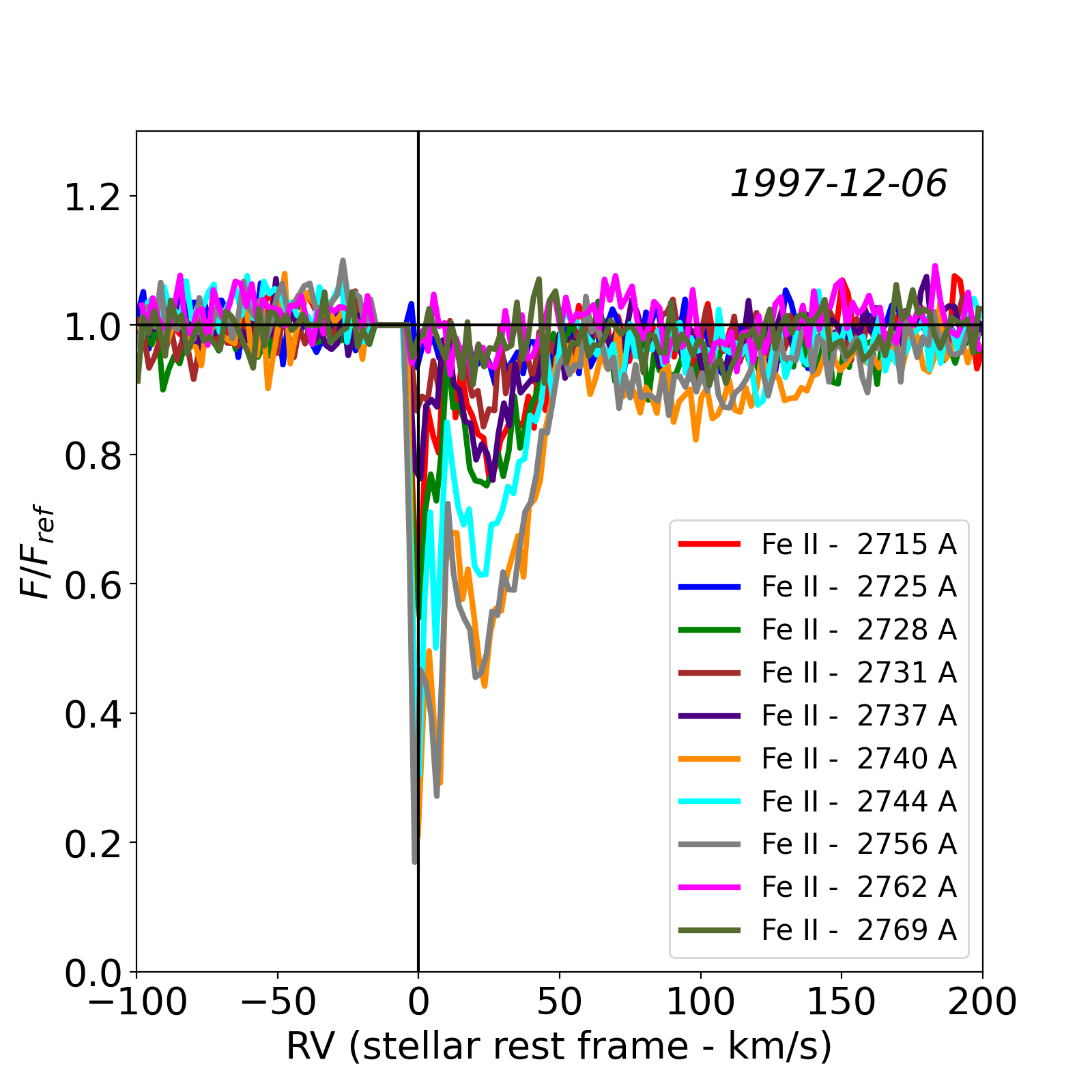}
 \includegraphics[clip, trim = 0 0 0 50,scale = 0.4]{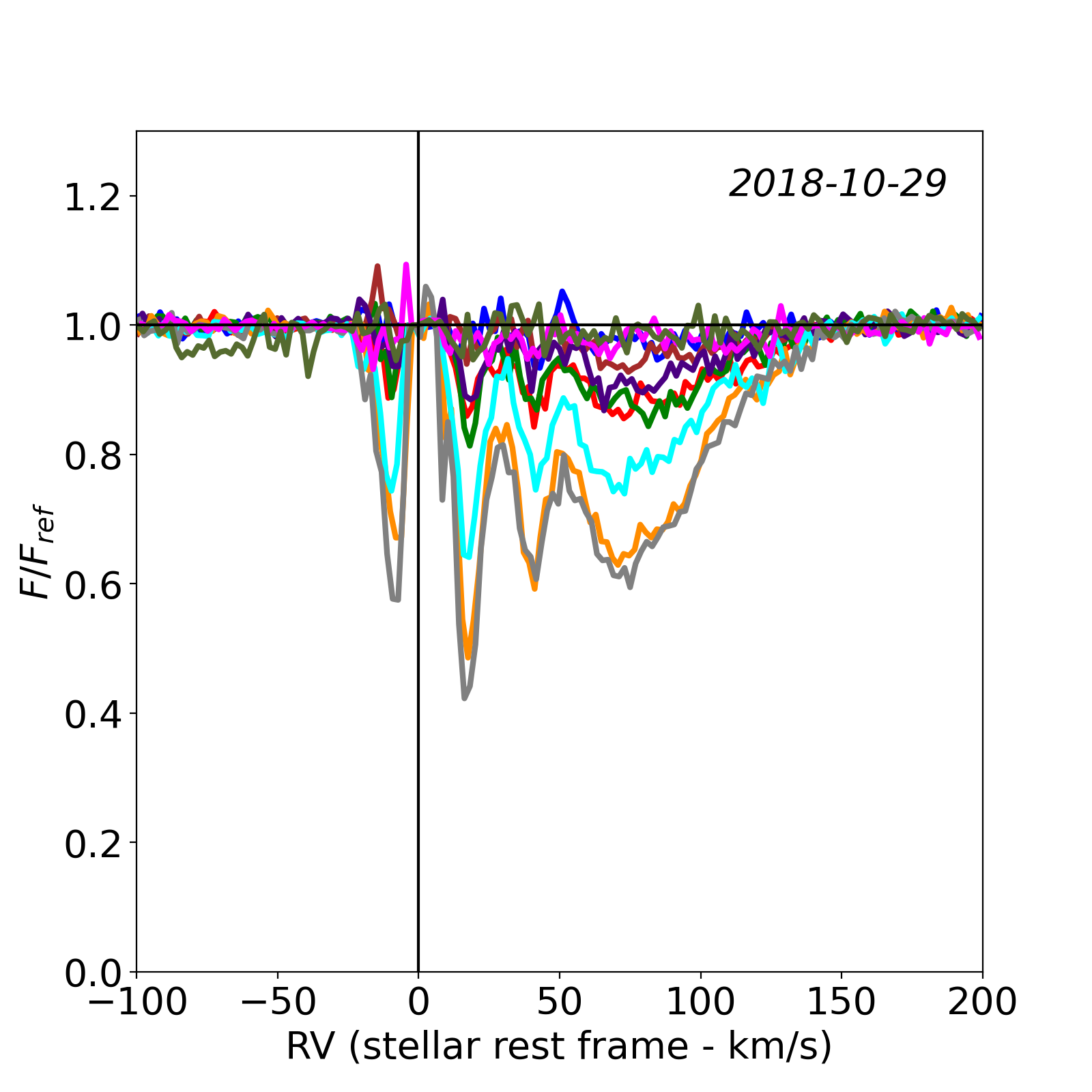}
      \caption[]{Cometary absorption spectra obtained as described in Sect. \ref{reconstruction of the reference} for the December 6, 1997 (top) and October 29, 2018 (bottom) observations, around ten \feii\ line around 2750\,\A. Low velocities absorptions (between $ -10$ and $+ 10 \ \si{km \ s^{-1}}$) may be attributed to comets or circumstellar variability. }
 \label{absorption}
\end{figure}

Cometary absorption spectra are obtained by dividing each individual observation by the comet-free spectrum previously calculated. This provides, for each observation, the absorption spectrum of the transiting gas: this spectrum is equal to~one in spectral ranges where the transiting comets are transparent, and tends to zero in the presence of an optically thick tail. These spectra show the striking correlation between the absorption features observed in various \feii\ lines, within the same observations. For example, Fig.~\ref{absorption} shows, for two separate visits, the superposition of their cometary absorption spectra around ten different \feii\ lines, clearly revealing the transit of numerous comets within each of the observations.\\

The analysis developed in the next sections is focussed on the 'absorption depth' and 'average absorption depth' of exocomets. These two quantities are defined as follows: 

\begin{itemize}
    \item For a given line ($l$, $u$) (where $l$ and $u$ denote the lower and upper levels of the transition, respectively) and some given radial velocity $v$, the absorption depth of exocomets is defined as the fraction of the stellar flux absorbed by exocomets at the velocity $v$ in the line ($l$, $u$) in a given observation. If we denote the cometary absorption spectrum computed previously by $\tilde{F}_{lu}(v),$  this absorption depth writes as 
    $$
    {\rm abs}_{lu}(v) = 1 - \tilde{F}_{lu}(v).
    $$    
    The absorption depths thus range from~zero in the absence of transiting comet to one when an optically thick cloud eclipses 100\% of the stellar surface. The error bars on the measured absorption depths are estimated by accumulating all the errors on each of the used HST spectra as tabulated by the {\tt calstis} pipeline.\\
    \item Then, to improve the signal-to-noise ratio (S/N) of our absorption measurements, the exocomet absorption depth can be averaged over an extended radial velocity range. For some radial velocities $v_1 < v_2$ and some line $(l,u)$, we thus introduce the mean absorption depth of exocomets, which is defined as the average value of ${\rm abs}_{lu}(v)$ over the RV domain $[v_1, v_2]$: 
    $$
    {\rm \overline{abs}}_{lu}(v_1, v_2) = \overline{  \, \Big\{   {\rm abs}_{lu}(v), \ \  v \in [v_1, v_2] \Big\} \,  }. \\
    $$
\end{itemize}

\section{Model description}
\label{First Model}

\subsection{Principle}

In order to interpret the absorption features due to the transit of exocomets, we used a modified version of the cometary model that \cite{Kiefer_2014} introduced to analyse the absorptions in the 3900\,\A\ \caii\ doublet of \bp. The gist of the model is to assume that the absorption depth (relative to the comet-free spectrum; see Sect. \ref{Cometary absorption}) observed at a radial velocity $v$ in a set of lines from a given chemical species is produced by a gaseous cloud of homogeneous column density, covering a fraction $\alpha(v)$ of the stellar disc surface. Under these assumptions, the absorption depth of a transiting comet is given, for any of the considered lines, by 
$$
{\rm abs}_{lu}(v) = \alpha(v) \cdot \big(1 - e^{-\tau_{lu}(v)} \big),
$$
with $l$ and $u$ being the lower and upper levels of the line (noted below $(l,u)$) and $\tau_{lu}(v)$ the optical thickness of the cometary cloud, which can be expressed as 
$$
\tau_{lu}(v) = \frac{hc}{4 \pi} \  B_{lu} \ {\rm N}_l \ \phi(v),
$$
where $B_{lu}$ is the Einstein coefficient for photon absorption of the considered line, N$_l$ the column density of the lower energy level of the transition and $\phi$ the radial velocity distribution within the cloud ($\int\phi(v)dv=1$). The Einstein coefficient $B_{lu}$ is linked to the oscillator strength of the transition $f_{lu}$ through 

\begin{equation}
    \frac{hc}{4 \pi} B_{lu}  = \frac{e^2}{4 \epsilon_0 m_e c} f_{lu} \lambda_{lu}. 
    \label{lien B f}
\end{equation}

Then, if we average the cometary absorption depth over a radial velocity range [$v_1; v_2$] where $\tau_{lu}(v)$ is assumed to be constant, we can re-write ${\rm abs}_{lu}$ and $\tau_{lu}$ as 

\begin{align}
\begin{aligned}
{\rm \overline{abs}}_{lu}(v_1, v_2) & = \overline{\alpha}(v_1, v_2) \cdot 
    \big(1 - e^{-\tau_{lu}(v_1, v_2)} \big),\\
\tau_{lu}(v_1, v_2) & = \frac{hc}{4 \pi} \ B_{lu} \ \frac{{\rm N}_l(v_1, v_2)}{v_2 - v_1},  
\label{equation modele général}
\end{aligned}
\end{align}
where ${\rm \overline{abs}}_{lu}(v_1, v_2)$ is the mean absorption depth in the radial velocity range [$v_1$; $v_2$] in the considered line, relative to the comet-free spectrum (see Sect. \ref{Cometary absorption}), $\overline{\alpha}(v_1 v_2)$ the mean size of the comet in this range, and N$_l(v_1, v_2)$ the column density of the considered species at the lower energy level of the transition with a radial velocity between $v_1$ and $v_2$. This set of equations is valid for any line of the considered chemical species.

\subsection{Two-parameter model}
\label{Two-parameters model}

Equation~\eqref{equation modele général} can be used to fit the mean absorption depth of a comet in a set of lines rising from the same excitation level of a given species, probing its column density ${\rm N}_l$. However, in the case where absorption features are detected in various lines rising from several excitation levels, one must find a way to link the various ${\rm N}_l$ of these levels in order to combine the information they provide and infer the global properties of the transiting comet. 

The easiest way of doing this is to consider the case where the excitation levels are sufficiently close in energy to be populated according to their multiplicity, that is, $N_l \propto g_l$. In particular, this hypothesis is valid at local thermodynamic equilibrium (LTE), when, for any pair of transition $(l_1, u_1)$ and $(l_2, u_2)$, we have
$$
|E_{l_1} - E_{l_2}| \ll k_B T,
$$
where T is the temperature of the absorbing medium, $E_{l_1}$, $E_{l_2}$ the lower level energies of the lines, and $k_B$ the Boltzmann constant.

In these conditions, we thus have $\tau_{lu}(v_1, v_2) \propto \lambda_{lu} g_l f_{lu}$ (Eqs.~\eqref{lien B f} and~\eqref{equation modele général}). We can therefore introduce a parameter $\beta$ defined such as 

$$
\tau_{lu}(v_1, v_2) = \beta(v_1, v_2) \cdot \frac{\lambda_{lu}}{\lambda_0} g_l f_{lu},
$$
where $\lambda_0$ is an arbitrary wavelength. For practicality, in the following $\lambda_0$ is chosen to be equal to the wavelength of the 2756\,\A\ \feii\ line; however, this choice has no impact on the results as it can be seen as a normalisation unit for $\beta$. Indeed, the parameter $\beta$ is a unitless quantity common to all the studied lines and close to the optical thickness when $g_l f_{lu} \sim 1$. $\beta$ is thus related to the column densities of the excitation levels giving rise to the observed lines through 

$$
\beta(v_1, v_2) g_l = \frac{e^2}{4 \epsilon_0 m_e c} \frac{{\rm N}_l(v_1, v_2)}{(v_2 - v_1)} \lambda_0.
$$

Using Eq. \eqref{cog eq}, we thus obtain the resulting curve of growth equation relating to the mean absorption depth within a radial velocity range to the line strength:

\begin{equation}
{\rm \overline{abs}}_{lu} = \overline{\alpha} \cdot \left(    1 - \exp\left(-\beta \frac{\lambda_{lu}}{\lambda_0} g_lf_{lu}\right)
\right).
\label{cog eq}  \end{equation}

We note here that ${\rm \overline{abs}}_{lu}$, $\overline{\alpha}$ and $\beta$ depend on the radial velocity range $[v_1; v_2]$ where the absorption depths are measured.
Thus, to use this model, one must first measure the mean absorption depths of a transiting comet in a set of lines with various absorption strengths ($g_l f_{lu}$), in a fixed RV range, and then fit the curve of growth model to these measured absorptions using Eq.~\eqref{cog eq} in order to estimate the values of $\overline{\alpha}$ and $\beta$ that best fit the observations. The  column densities ${\rm N}_l$ of the studied excitation levels can be then  derived from the value of $\beta$ using

$$
{\rm N}_l(v_1, v_2) = \frac{4 \epsilon_0 m_e c}{e^2 \lambda_0} (v_2 - v_1) \  \beta g_l.
$$

This two-parameter model was used by \cite{Kiefer_2014} to determine $\overline{\alpha}$, the typical size of the gaseous clouds, of almost 500~comets transiting \bp\ detected in the \caii\ doublet. However, in this case, the model was fit to only two absorption measurements (one for each line of the doublet); it was thus not possible to check that it was realistic. To show that the model is correct and that the measured absorptions really follow a curve of growth as given by Eq.~\eqref{cog eq}, a larger number of lines with various oscillator strengths is needed. The many ($> 10$) \feii\ lines observed with STIS, which show clear cometary absorptions, are thus a very useful tool to test the curve of growth model (Sect.~\ref{validation 2 parameters}).

\subsection{Thermodynamical equilibrium: A three-parameter model}

In the case where the cometary absorptions are detected in lines arising from a large variety of excitation levels, the hypothesis $N_l \propto g_l$ may no longer be a good approximation. For instance, very excited levels may be under-populated compared to ground states. To model the relative abundances of the considered energy levels, another possibility is to assume that the absorbing cloud is at LTE, with a typical temperature of $T(v_1, v_2)$ (where, again, [$v_1$; $v_2$] is the considered radial velocity range). This implies

$$
N_l(v_1, v_2) \propto g_l e^{-E_l/k_B T(v_1, v_2)}.
$$

We note that this new assumption requires a sufficiently high electronic density to impose a collisional regime.
At LTE, Eqs.~\eqref{lien B f} and~\eqref{equation modele général} now yield $\tau_{lu}(v_1, v_2) \propto \lambda_{lu} f_{lu} g_l e^{-E_l/k_B T(v_1, v_2)}$. This allows the introduction of a new parameter $\gamma(v_1, v_2)$, linking the optical thickness to the line strength through
$$
\tau_{lu}(v_1, v_2) = \gamma(v_1, v_2) \cdot \frac{\lambda_{lu}}{\lambda_0} g_l f_{lu} e^{-E_l/k_B T(v_1, v_2)}.
$$

Once again, $\gamma$ is a unitless parameter common to all studied lines, close to the optical thickness when $g_l f_{lu} e^{-E_l/k_B T}$$\sim$1, and which depends on the radial velocity range where the mean absorption depths are measured. Thanks to Eqs.~\eqref{lien B f} and~\eqref{equation modele général}, it can be linked to the abundance of the considered excitation levels within $[v_1, v_2]$, through: 
$$
\gamma(v_1, v_2) \ g_l e^{-E_l/k_B T(v_1, v_2)} = \frac{e^2}{4 \epsilon_0 m_e c} \frac{{\rm N}_l(v_1, v_2)}{(v_2 - v_1)} \lambda_0.
$$

Summing the previous equation over all the excitation levels of the considered species, one then deduces 
$$
\gamma(v_1, v_2) \ Z\Big(T(v_1, v_2)\Big) = \frac{e^2}{4 \epsilon_0 m_e c} \frac{{\rm N}_{\rm tot}(v_1, v_2)}{(v_2 - v_1)} \lambda_0,
$$
where ${\rm N}_{\rm tot}(v_1, v_2)$ reflects the total column density of the species with radial velocity between $v_1$ and $v_2$, and Z is defined by 
$$
Z(T) = \displaystyle \sum_{i=0}^{+ \infty} g_i \ e^{-E_i/k_B T}.
$$

We therefore end up with a three-parameter curve of growth model that links the mean absorption depths of a transiting cloud in a set of lines of a given species, measured on a fixed radial velocity range, to the mean size ($\overline{\alpha}$), temperature ($T$), and total column density (via the $\gamma$ parameter) of the absorbing cloud, through
\begin{equation}
{\rm \overline{abs}}_{lu} = \overline{\alpha} \cdot \left(    1 - \exp\left(-\gamma \frac{\lambda_{lu}}{\lambda_0} g_l f_{lu}  e^{-E_l/k_B T}\right)
\right),
\label{cog eq 3}    
\end{equation}
where, again, ${\rm \overline{abs}}_{lu}$, $\overline{\alpha}$, $\gamma,$ and $T$ depend on the radial velocity range $[v_1; v_2]$ where the mean absorption depths are measured. Once these three quantities have been constrained by a curve of growth model fit to the measured absorption depths, one can estimate the total column density of the studied species through
$$
{\rm N}_{\rm tot}(v_1, v_2) = \frac{4 \epsilon_0 m_e c}{e^2 \lambda_0} (v_2 - v_1) \  \gamma \ Z\big(T\big).
$$

We note that this model is valid under the assumption of column density and temperature homogeneity of the absorbing gas over the considered radial velocity range. This hypothesis can be checked by fitting the curve of growth model (Eq. \eqref{cog eq 3}) to the cometary absorptions, as done in Sect.~\ref{Temperature and electron density}. If the measured absorptions do indeed follow the model, then the fitted parameters ($\overline{\alpha}$, $\gamma$, $T$) may represent valuable estimates of the typical properties of the transiting cometary cloud.

Finally, the two-parameter model described in Sect.~\ref{Two-parameters model} can be seen as a simplified case of the three-parameter model, where the temperature $T$ is taken to be much higher than the energy difference between the excitation level of the considered lines. Indeed, in such a case, $e^{-E_l/k_B T}$ is constant over the set of lines, thus yielding $N_l(v_1, v_2) \propto g_l$ (Sect. \ref{Two-parameters model}). The $\beta$ parameter introduced in the previous section is then linked to $\gamma$ through $$
\beta \simeq \gamma e^{-E_l/k_BT},
$$
where this equation holds for any of the considered energy level $l$. In particular, this is the case when there is only a single excitation level for the observed set of lines, as for the \caii\ doublet at 3933 and 3369\,\AA.

\section{Validation of the two-parameter model}
\label{validation 2 parameters}

In order to validate our different models, we focus on two separate events in the following. The first one is the transit that occurred on December 6, 1997, between +12 and +40 km/s (see Fig. \ref{absorption}). Its main advantage is that it was observed in a great variety of \feii\ lines (in particular, in the three main series at 2400, 2600, and 2750 \A), which will be useful for validating the three-parameter model in the next section. The second transit is the one from the October 29, 2018 observation, between +50 and +150 km/s (see Fig. \ref{absorption}), which benefits from a very good S/N (about 100) due to the long duration of the exposure. Another advantage of these transits is their high radial velocities, making them unaffected by the circumstellar absorption near 0~km/s (which reduces the S/N and may show long-term variability, \cite{Kiefer2019}).

The two-parameter curve of growth model was validated using the \feii\ line series around between 2715 and 2769\,\A\ arising from the a$^4$D \feii\ term. Indeed, the four levels that make up this state share close excitation energies (corresponding to temperature between 11448 and 12771 $\si{K}$; see Tables \ref{list lines} and \ref{list levels}), allowing us to make the hypothesis - which we verify below - that their relative abundances are proportional to their multiplicity.

\begin{figure}[h]
 \centering
  \includegraphics[clip, trim = 0 0 40 60, scale = 0.36]{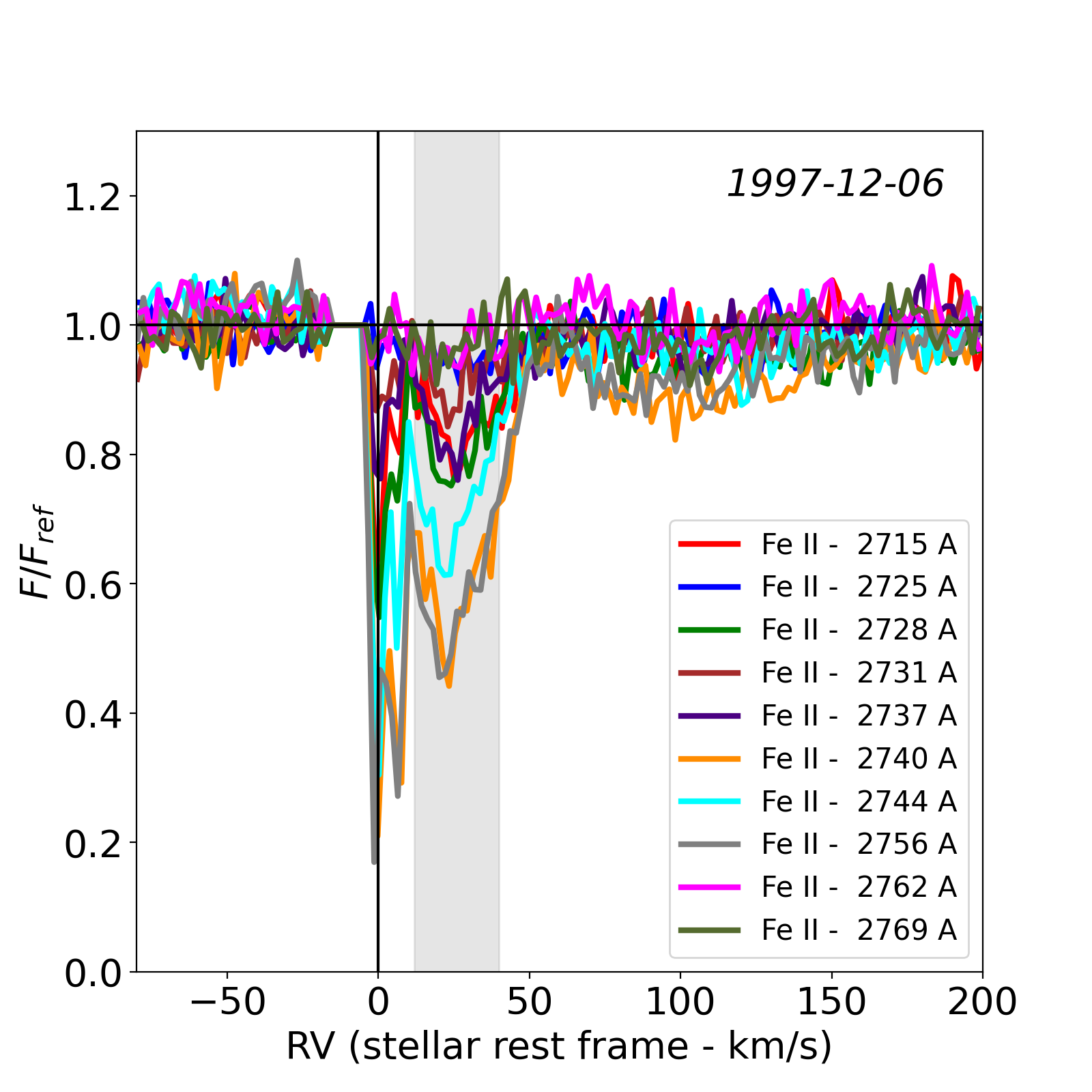}  \includegraphics[clip, trim = 20 10 0 50, scale = 0.4]{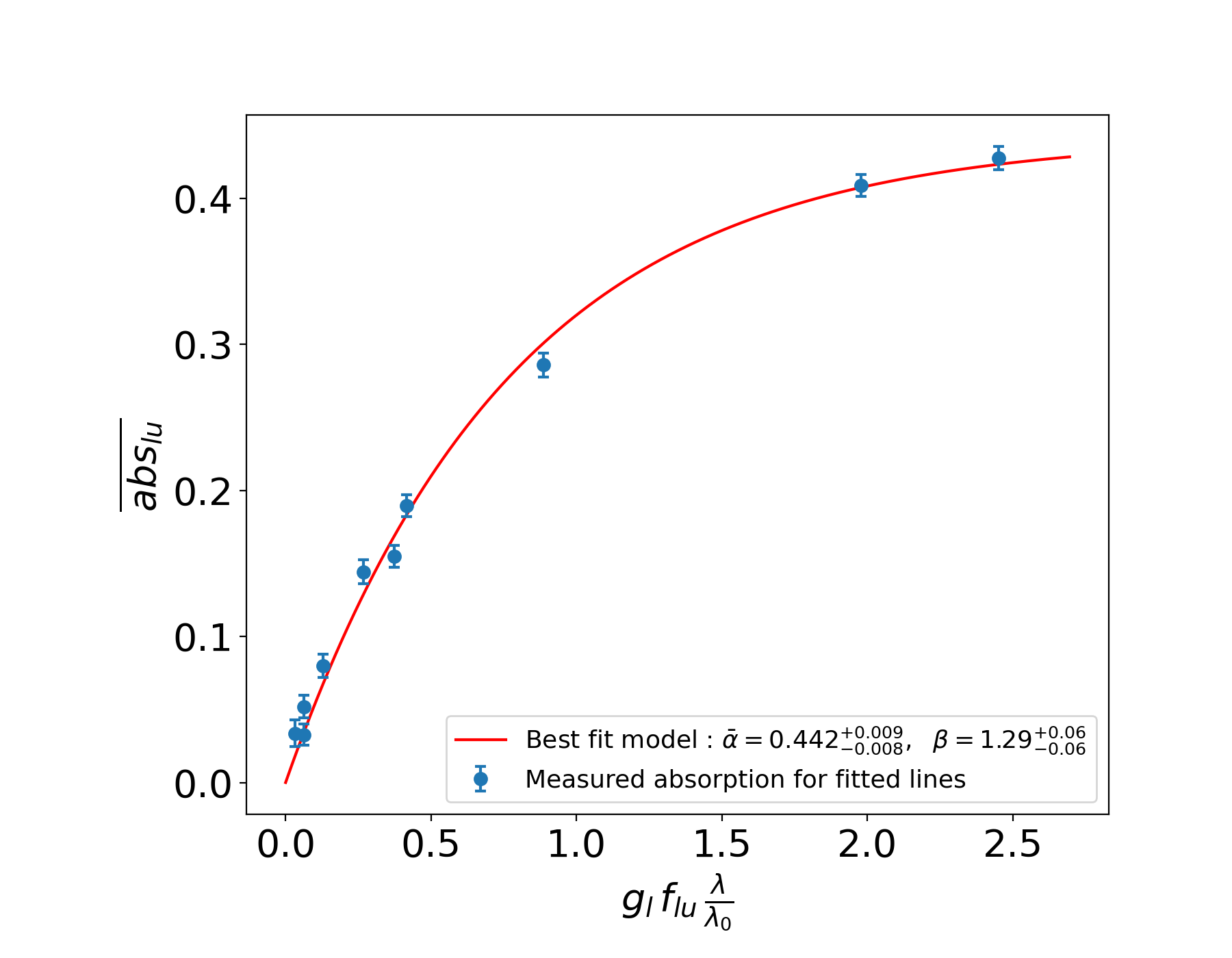}

      \caption[]{Analysis of the December 6, 1997 comet. \textbf{Top}: Cometary absorption spectrum of December 6, 1997 observation in the a$^4$D \feii\ lines around 2750\,\AA . The grey zone represents the radial velocity range [+12; +40]~km/s over which the cometary absorptions were measured. \textbf{Bottom}: Curve of growth of the same \feii\ lines, showing the mean absorption depths measured in the [+12; +40]~km/s range (blue dots), with one-sigma error bars computed from the STIS pipeline tabulated error bars and the two-parameter fitted model (red line) as a function of the theoretical line strength ($g_l f_{lu} \frac{\lambda_{lu}}{\lambda_0}$). 
      }
 \label{COG o4g infinite T}
\end{figure}

Figure~\ref{COG o4g infinite T} shows a first example of a curve of growth fit (lower panel) for the redshifted comet observed in the December 6, 1997 data (upper panel). The mean absorption depths were measured in the [+12; +40]~km/s radial velocity range, corresponding to the main domain where this comet is visible. These absorption measurements were then fit with the two-parameter curve of growth model, using a Markov chain Monte Carlo (MCMC) algorithm. Remarkably, the ten absorption measurements follow the two-parameter curve of growth model very well (reduced $\rchi^2$ of 2.7), constraining the average size of the transiting comet to $44 \pm 1 \%$ of the stellar surface.

\begin{figure}[h!]
 \centering
  \includegraphics[clip, trim = 0 0 40 60, scale = 0.36]{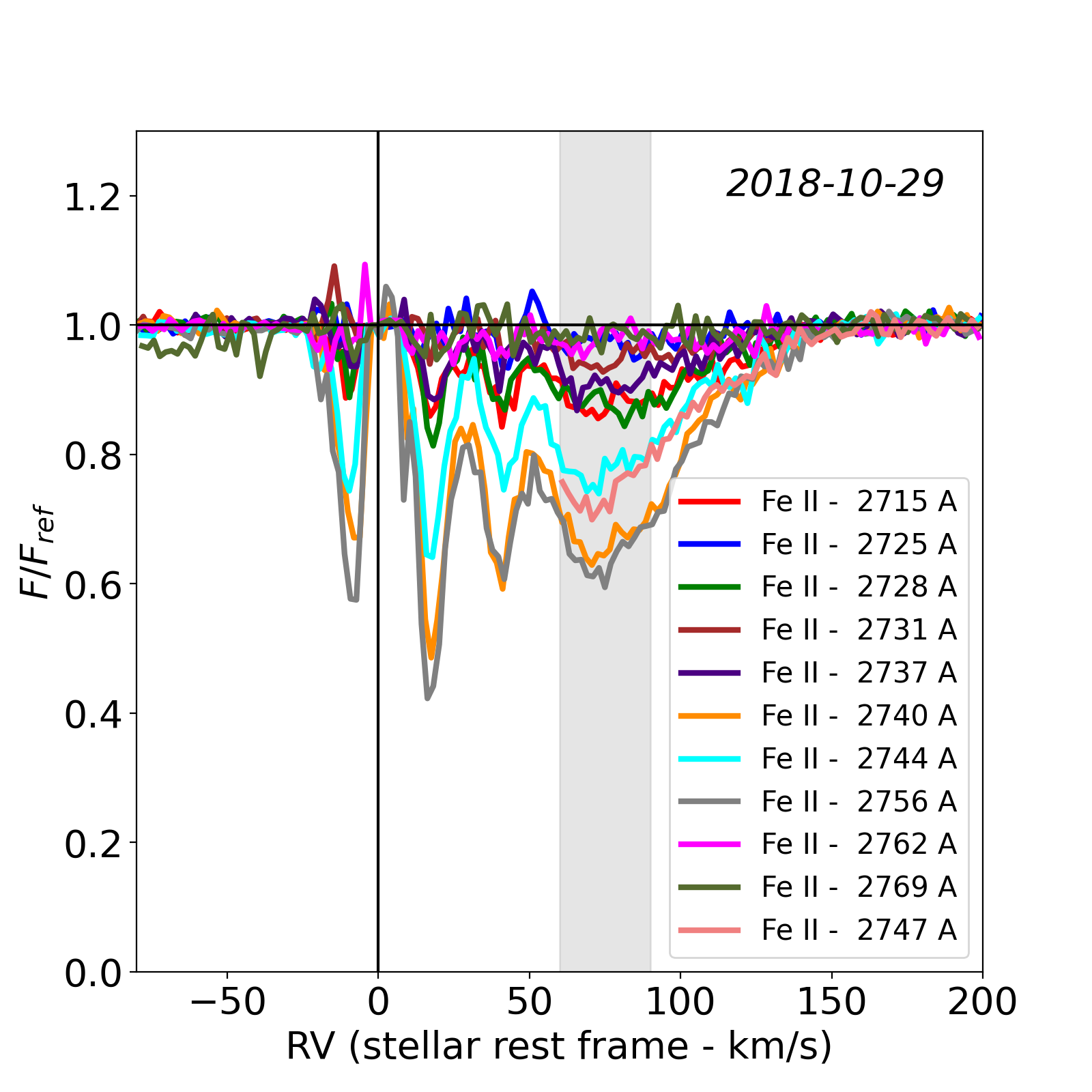}  \includegraphics[clip, trim = 20 10 0 50, scale = 0.4]{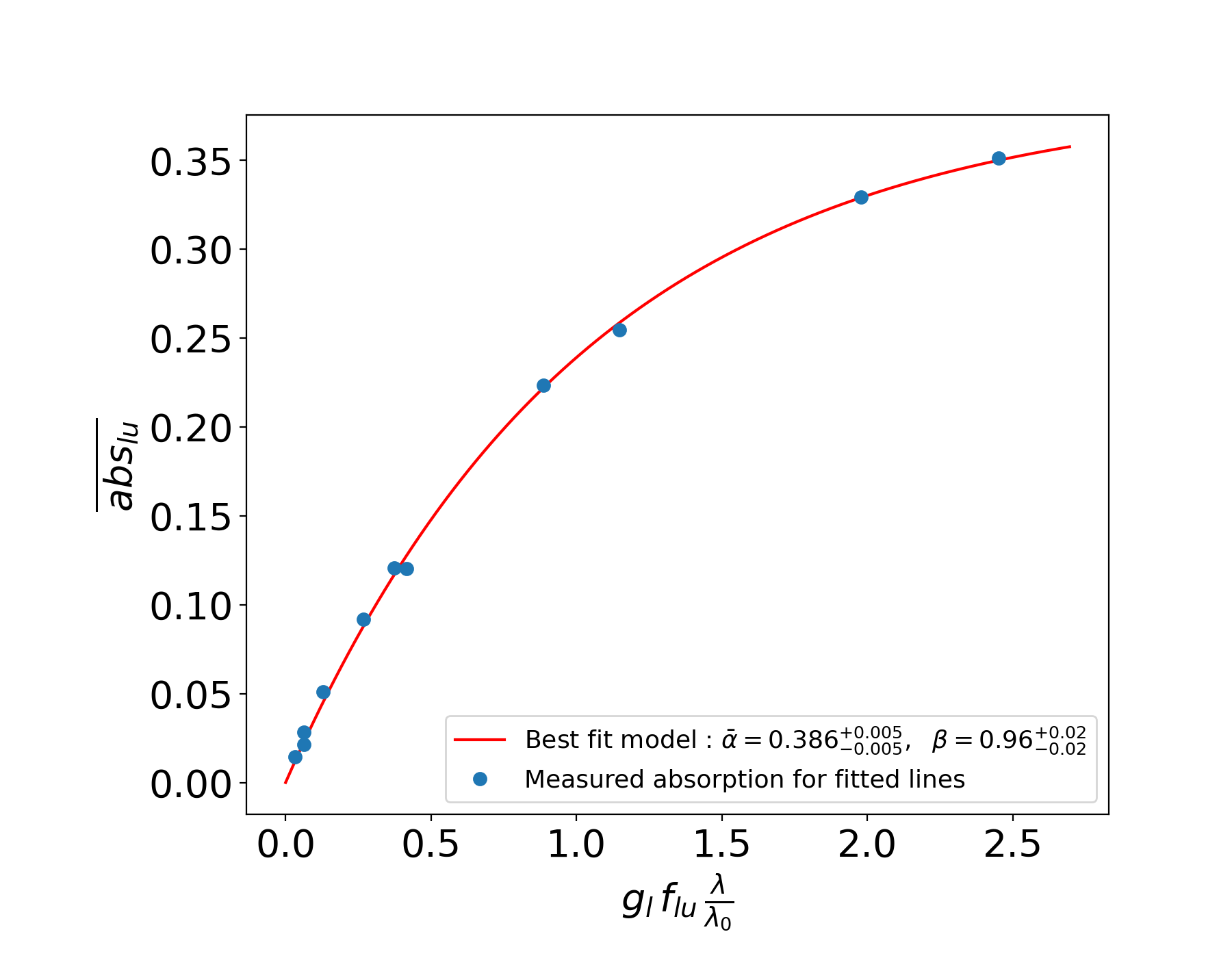}
      \caption[]{Analysis of the October 29, 2018 comet. \textbf{Top}: Cometary absorption spectrum of October 29, 2018 observation in a$^4$D \feii\ lines around 2750\,\AA. The grey zone represents the radial velocity range [+60; +90]~km/s, which was used to measure the cometary absorptions. 
      \textbf{Bottom}: Curve of growth of the same \feii\ lines, showing the mean absorption depths measured in the [+60; +90]~km/s range (blue dots) and the two-parameter fitted model (red line) as a function of the line absorbing strength. The uncertainties on the absorption measurements are of the order of the marker's size, thanks to the very good S/N ($\sim 100)$ of the October 29, 2018 observation.}  
 \label{COG odt infinite T}
\end{figure}

Another example is given in Fig.~\ref{COG odt infinite T} for the October 29, 2018 event, where the comet mean absorption depths are measured from +60 to +90 km/s (the radial velocity range where it is the strongest). For this comet, we also took into account the reddest line of the 2747 \A\ doublet (see Fig. \ref{Fe II lines 2750}). Again, the two-parameter model fits the 11 measured absorptions strikingly well (reduced $\rchi^2$ of 1.9), and constrains the mean size of the comet to $39 \pm 1 \%$ of the stellar disc (within this velocity range).

These two examples clearly show that the variable features observed in the \feii\ lines of \bp \ are explained well by the transit of homogeneous cometary clouds, covering large but limited fractions of the stellar disc, and in which the sub-levels of the a$^4$D \feii\ term have relative abundances proportional to their multiplicity. This remarkable agreement between the measured absorptions and the two-parameter curve of growth model also validates the results of \cite{Kiefer_2014}, who applied this model to the \caii\ doublet at $\sim$3900\,\A\ to identify two families of exocomets in the \bp\ system. 

In addition, if we assume that the transiting cloud is at LTE (which is probably a reasonable assumption, as shown in Sect.~\ref{Temperature and electron density}) and use the three-parameter model to fit the measured absorption depths (Eq.~\eqref{cog eq 3}), we can put lower limits on the gas temperature around $5\,000\,\si{K}$ for the December 6, 1997 comet, and $10\,000\,\si{K}$ for the October 29, 2018 comet, at a 95 \% confidence level. This is due to the fact that the lines rising from the most excited levels of the a$^4$D term ($E_l$\,$\sim$\,$12\,700\,\si{K}$; see Table \ref{list levels}) do not show significantly lower absorption than the lines rising from the less excited levels ($E_l$\,$\sim$\,$11\,400\,\si{K}$): the gas temperature must thus be well above the energy difference between all these states, that is, $\gg$\,$1\,000\,\si{K}$.

Lower limits (with a $2 \sigma$ confidence) can also be set on the total \feii\ column density within the transiting clouds in the considered radial velocity ranges, at about $9.2 \times 10^{14} \ \si{cm^{-2}}$ (December 6, 1997) and $8.2 \times 10^{14} \ \si{cm^{-2}}$ (October 29, 2018), which would be reached for temperatures between $11\,000$ and $15\,000\,\si{K}$.

The very high temperatures hinted in the tails of these two comets (similar to or above the effective temperature of \bp) are consistent with the results that were obtained by \cite{Mouillet_1995} using \caii\ triplet lines arising from the 3D metastable level. This shows that such objects are experiencing very strong compression, which 
releases heat \citep[see][]{Beust_1993}. This compression is also likely responsible for the production of the observed  highly ionised species such as \aliii\ or \siiv.


\section{Validation of the three-parameter model}
\label{Temperature and electron density}

\subsection{Temperature measurement}
\label{Temperature measurement}

\begin{figure}[h!]
 \centering
 \includegraphics[clip, trim = 20 20 0 70, scale = 0.36]{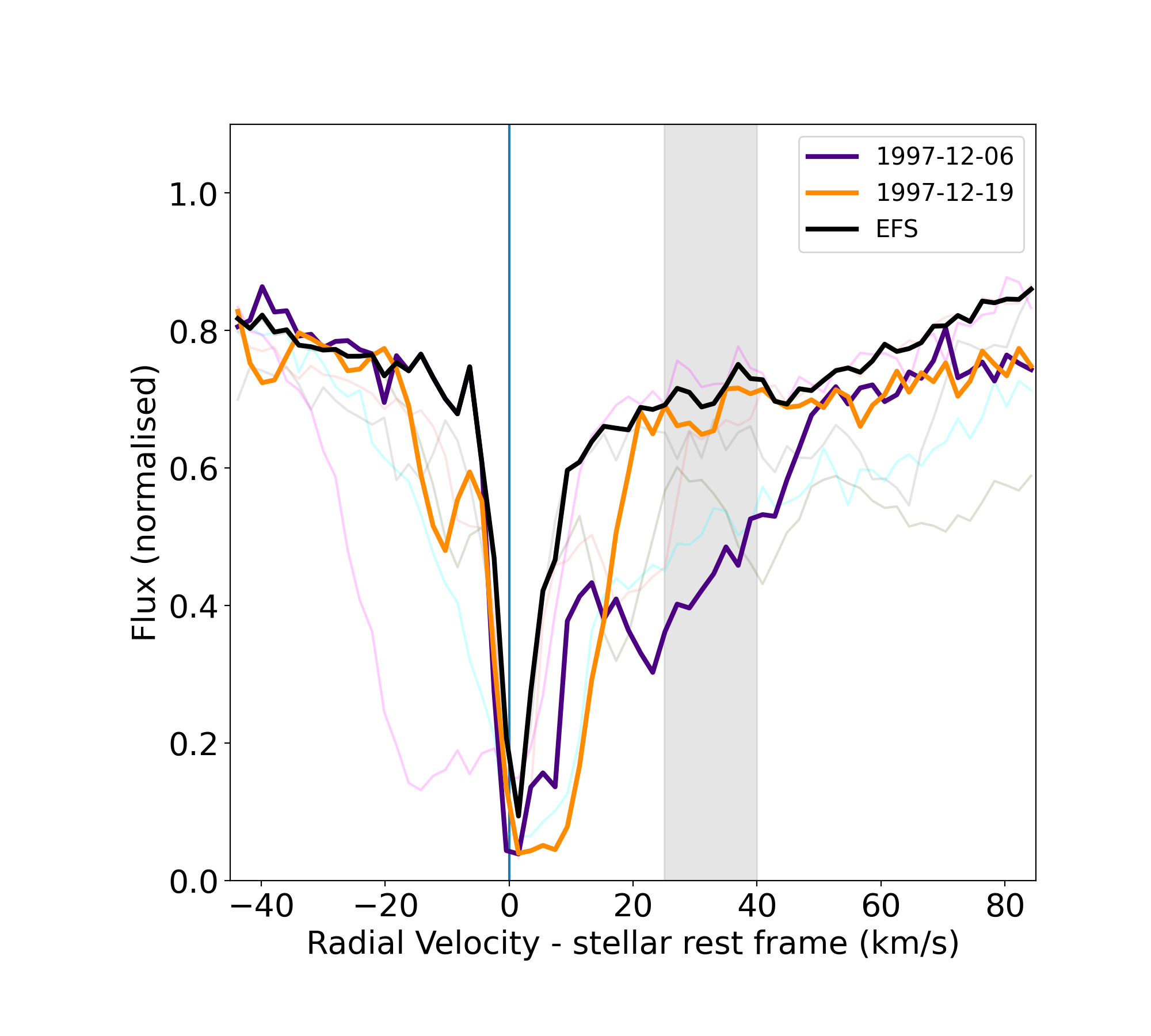} \includegraphics[clip, trim = 20 20 0 70, scale = 0.36]{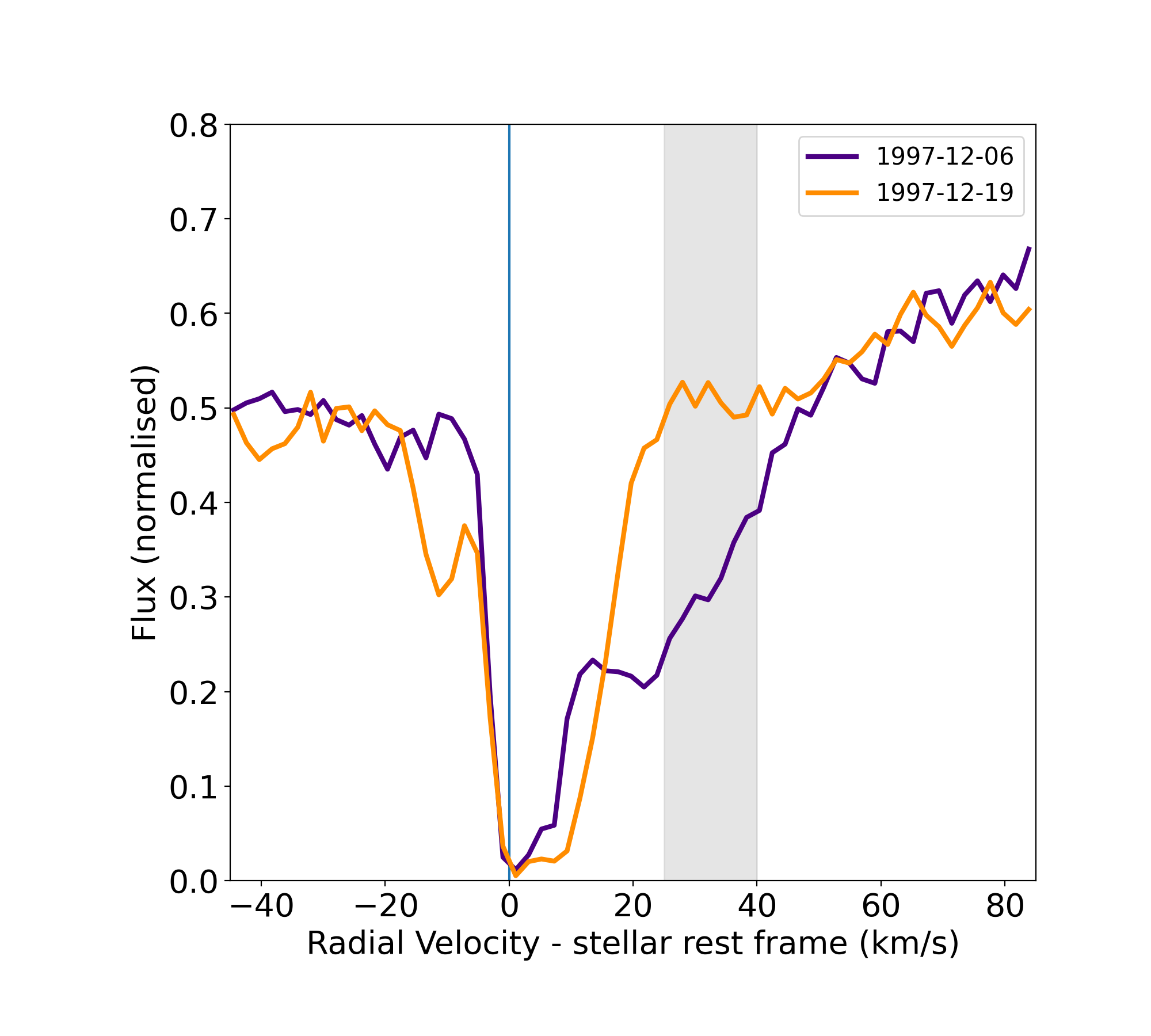}
      \caption[]{\bp\ spectra around two \feii\ lines. \textbf{Top}: Zoomed-in view of the 2740 \A\ \feii\ line  ($E_l=7\,955\,{\rm cm}^{-1} $), emphasising the two 1997 observations. The December 19 spectrum is close to the stellar spectrum in the [+25; +40]~km/s range (grey zone), while the December 6 observation shows a clear cometary transit at these radial velocities. \\
      \textbf{Bottom}: Zoomed-in view of the 2607 \A\ \feii\ line} ($E_l=668\,{\rm cm}^{-1}  $), showing the only two available spectra for this line.
 \label{2740_1997_obs} 
\end{figure}

The analysis in Sect.~\ref{validation 2 parameters} yielded striking results about the shape and thermodynamical conditions in the transiting cometary clouds of \bp. However, only lower limits could be set on their temperature and column density, as the energy difference between the different sub-levels of the \feii\ a$^4$D term (which gives rise the line series at 2750 \A; see Tables \ref{list lines} and \ref{list levels}) was too low to probe the energy distribution of Fe$^+$. Luckily, the two 1997 observations also cover the 2300-2600\,\A\ wavelength range, where a large number of \feii\ lines rising from a wider range of excitation levels are present (see Table~\ref{list lines}). In addition to the ten~previously studied lines, we thus included in our analysis 31~new lines with wavelengths between 2328\,\AA\ and 2629\,\AA , and with lower level energies from 0 to 12700~K (0 - 8000 $\si{cm^{-1}}$). For the lines arising from low energy levels ($\leq 4500 \ \si{K}$), we limited ourselves to the ones with average values of the absorbing strength ($0.08 < g_l f_{lu} < 0.7$), avoiding both the weakest transitions ($g_l f_{lu} < 0.08$), for which the S/N on the absorption depth is weak, and the strongest ones ($g_l f_{lu} > 0.7$), which showed abnormally strong absorption depths (see the discussion in Sect.~\ref{Discussion}).
However, wavelengths below 2600\,\AA\ are only covered by the two 1997 observations. This prevents us from building a proper reference spectrum at these wavelength (as we did in Sect. \ref{reconstruction of the reference} for the 2750 \A\ lines), as there are wide ranges of radial velocities where both 1997 spectra seem affected by comets (see Table \ref{RV_ranges_comet}) and where, therefore, it is impossible to recover the unnoculted stellar flux. To exploit these lines and test our three-parameter curve of growth model, we thus chose to focus on the redshifted comet observed on December 6, 1997 (Fig.~\ref{COG o4g infinite T}) and limited our study to the [+25; +40]~km/s radial velocity range. Indeed, in this narrower velocity range, the other 1997 observation taken 13~days later seems to be little affected by comets (Fig. \ref{2740_1997_obs}), providing a sufficiently reliable estimate of the unnoculted \bp\ spectrum and enabling us to retrieve the absorption depth of the December 6, 1997 comet in the \feii\ lines between 2300\,\AA\ and 2600\,\A. Here, for the sake of consistency, we also used the same December 19, 1997 observation as a reference spectrum for the 2750 \A\ lines, despite the fact that other observations are available. This choice, however, has very little incidence on our measurements; the results would be very similar if we had used the 'comet-free' spectrum as a reference for these lines, as we did in the previous section.

\begin{figure*}[t]
    \centering    
    \includegraphics[clip, scale = 0.4, trim = 80 20 50 20]{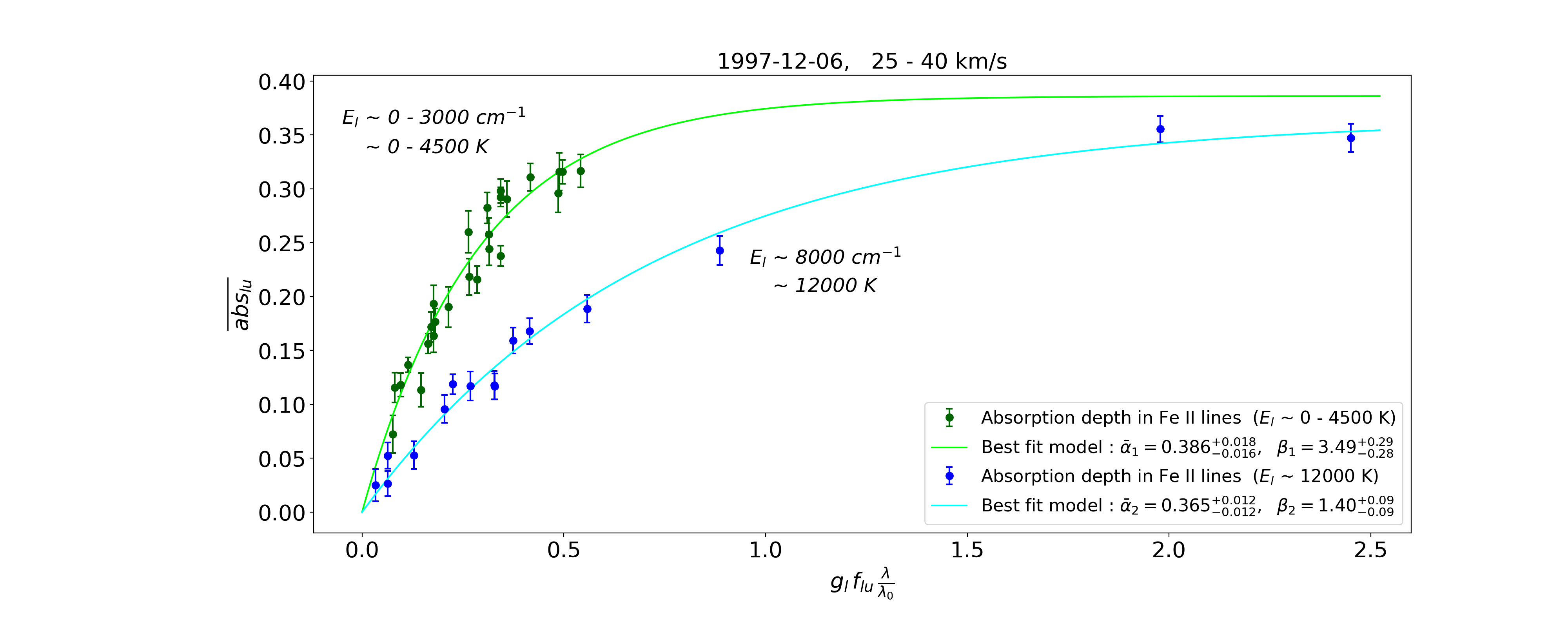}
    \caption{Curve of growth of 26 \feii\ lines rising from low energy levels ($0 - 4500 \ \si{K}$, green) and 15 lines rising from high energy levels ($\sim 12000 \ \si{K}$, blue) for the December 6, 1997 comet in the $[+25, +40]$ km/s RV range. While both sets of lines provide very similar values for the average size of the absorbing cloud ($\sim 38$ \%), the lines rising from the ground terms saturate a lot quicker.}
    \label{COG_separate_levels}
\end{figure*}

First, we started by separately fitting the mean absorption depths of the December 6, 1997 comet in the lines rising from low-energy levels  ($E_l$\,$\la$\,$4\,500\,\si{K}$; i.e. $E_l$\,$\la$\,$3\,000\,\si{cm^{-1}}$) and higher energy levels ($E_l$\,$\sim$\,$12\,000\,\si{K}$\,$\sim$\,$8\,000\,\si{cm^{-1}}$), 
using the two-parameters curve of growth model tested above. The results are provided in Fig. \ref{COG_separate_levels}; we note that the two sets of lines constrain the size of the absorbing cloud to very close values ($38.6 \pm 1.8$ \% for the ground levels lines, $36.5 \pm 1.2$ \% for the metastable lines), hinting that the absorptions occurring in both sets are produced by the same gaseous component. We note, however, that the lines rising from ground \feii\ levels are more easily saturated than the lines rising from excited levels ($\beta \sim 3.5$ for the first, $\beta = 1.4$ for the last): this reveals the significant under-population of the excited levels due to a finite temperature in the absorbing medium.

\begin{figure}[h!]
 \centering
  \includegraphics[clip, trim = 30 10 0 60, scale = 0.42]{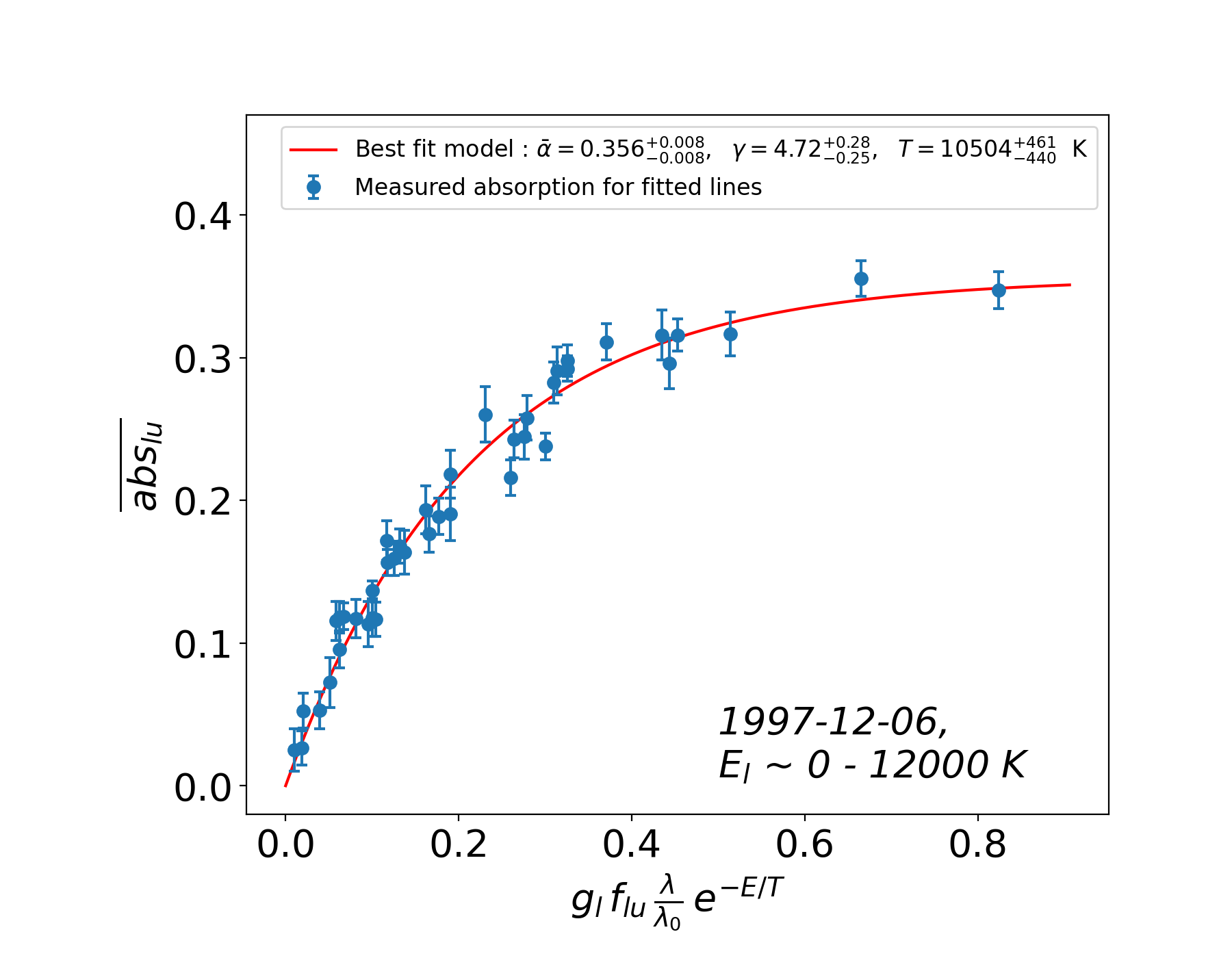}  
      \caption[]{Curve of growth of 41 \feii\ lines rising from various energy levels (from 0 to 12\,700\,$\si{K}$). The cometary absorption depths (blue dots) were measured in the [+25; +40]\,km/s velocity range in the December 6, 1997 observations.}

 \label{o4g_temperature}
\end{figure}

We then fitted the 41 \feii\ lines all together, with the addition of a new free parameter: the temperature. Despite the very high number of measurements, three free parameters seem to be enough to fit the data all together. Indeed, our curve of growth model matches the measured absorption depths remarkably well ($\rchi^2_r = 1.7$), allowing us to constrain the temperature of the absorbing medium at $10\,500 \pm 500\,\si{K}$ and its average size at $36 \pm 1 \%$ of the stellar surface (Fig.~\ref{o4g_temperature}). The total \feii\ column density in the studied radial velocity range (i.e. between +25 and +40 km/s) was thus estimated to be $6.7 \pm 0.4 \times 10^{14} \si{cm^{-2}}$.

In order to better visualise the temperature of the absorbing gas, it is also possible to build an excitation diagram of Fe$^+$ within the comet. Indeed, the model Eq.~\eqref{cog eq 3} can be re-written as 
$$e^{-E_l/k_BT} = \frac{- \ln(1-({\rm \overline{abs}}_{lu} \big/ \overline{\alpha}))}{\gamma (\lambda_{lu}/\lambda_0) g_l f_{lu} },
$$

yielding 
\begin{equation}
    \frac{-E_l}{k_BT \ln(10)}  = \ \log \Big(\frac{\text{N}_l}{g_l \text{N}_{\text{\rm tot}}} Z(T) \Big) =  \ \log \Big( \frac{- \ln(1- ({\rm \overline{abs}}_{lu}/\overline{\alpha}))}{\gamma (\lambda_{lu}/\lambda_0) g_l f_{lu}} \Big).
    \label{equation diagram excitation}
\end{equation}

\begin{figure}[h!]
 \centering
  \includegraphics[clip, trim = 15 25 0 60, scale = 0.42]{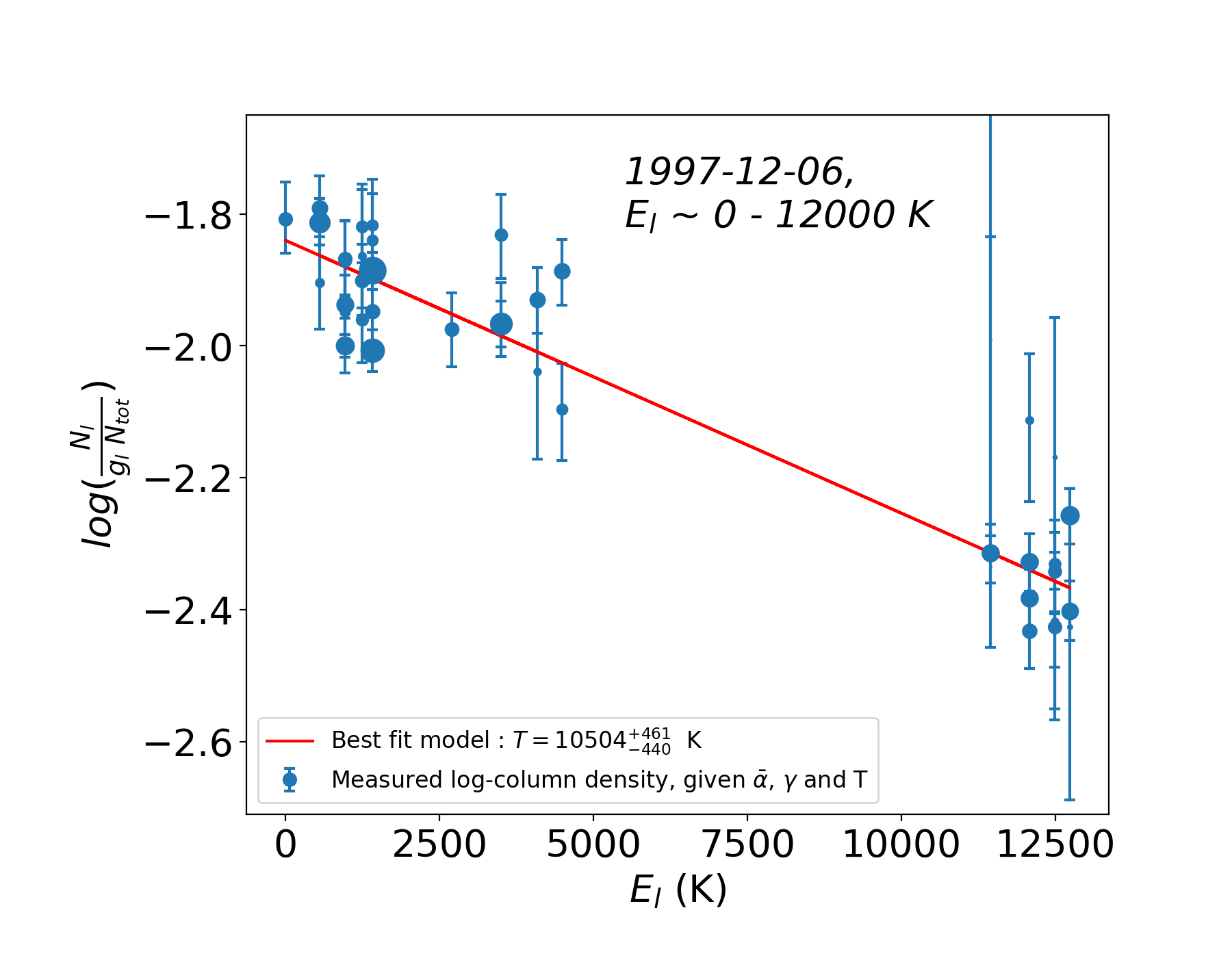}  
      \caption[]{Excitation diagram of \feii\ within the December 6, 1997 comet, obtained using absorption measurements in the [+25; +40]\,km/s radial velocity range}. The y-axis represents the log-relative abundance of the studied \feii\ energy levels, within an additive constant (see Eq. \eqref{equation diagram excitation}), while the x-axis indicates the energy of these levels. Each point thus represents an abundance estimate of an energy level $l$, based on the measurement of the absorption depth in a given line $(l,u)$. As some lines rise from the same energy level, some of these levels benefit from several independent abundance measurements. The marker sizes are inversely proportional to the vertical uncertainties.
 \label{excitation_diagram}
\end{figure}

The slope of $\log(N_l/g_l)$ expressed as a function of $E_l$ is therefore inversely proportional to the temperature $T$. On the other hand, knowing $\overline{\alpha}$, which characterises the typical size of the absorbing cloud independently of the line considered, we can derive an estimate of $\log(N_l/g_l)$ from each absorption measurement ${\rm abs}_{lu}$ in any spectral line arising from the energy level $l$ to within one additive constant depending on $\gamma$, $Z(T),$ and ${\rm N}_{\rm tot}$. Therefore, the temperature of the transiting cloud can be directly estimated through the variation of ${\rm N}_l/g_l$ (the relative abundance of the various Fe$^+$ excitation levels, normalised by their multiplicity)  as a function of $E_l$. This is illustrated in Fig.~\ref{excitation_diagram}, where we can clearly see that the a$^4$D energy levels ($E_l \sim 12\,000 \si{K}$) are significantly under-populated when compared to the ground levels due to the finite temperature ($\sim 10\,500 \si{K}$) within the transiting gas.

We can finally extrapolate this temperature to the whole velocity range of the December 6, 1997 comet (i.e. between +12 and + 40 km/s) in order to estimate its total column density (which was found in Sect. \ref{validation 2 parameters} to be greater than $9.2  \cdot 10^{14} \ \si{cm^{-2}}$). Indeed, fitting the 2750\,\A\ lines' absorption depths measured in this velocity range again with a Gaussian prior on the temperature of $10\,500 \pm 500 \, \si{K}$, one can deduce a total \feii\ column density in the transiting gas of $1.11 \pm 0.09 \times 10^{15} \, \si{cm^{-2}}$. Given the relative size of the comet ($\sim$$44\%$ of \bp\ surface, Fig. \ref{COG o4g infinite T}), the radius of \bp\ ($\sim$$1.8 \ R_\odot$) and the mass of an iron atom ($\sim$$9.4\times 10^{-23}\,\si{g}$), we can convert this column density into a mass of $\sim$\,$2.2 \times 10^{15} \, \si{g}$, which is equivalent to $0.28 \, \si{km^{3}}$ of pure iron (density $\simeq 8$). This tremendous mass shows the intense evaporation occurring at the surface of these cometary nuclei, hinting that they have very short lifetimes and may not survive to even a single periastron passage. This remark therefore indicates that a large fraction of the comets observed to transit \bp\ may originate from a recent break-up of a former massive body, as was suggested in \cite{Kiefer_2014} for the \emph{D} family of \bp\ comets.

\subsection{The electronic density}
\label{The electronic density}

Until now, our study was focussed on \feii\ lines arising from energy levels below 12\,000\,$\si{K}$, assuming that the transiting cloud is at LTE. Using the ChiantiPy Python module \citep{ChiantiPy_2012}, which allows us to compute the energy distribution of many ions at given temperature and electronic density ($n_e$), it can be shown that this hypothesis is easily verified, roughly requiring $n_e \geq 10^5\,\si{cm^{-3}}$. This allowed for an estimate of the absorbing gas temperature, assuming that the excitation temperature of the low-energy \feii\ levels (deduced from the excitation diagram, Fig.~\ref{excitation_diagram}) is equal to the kinetic temperature. 

In addition to the lines studied in Sects. \ref{validation 2 parameters} and \ref{Temperature and electron density}, cometary absorption is also detected in a few lines arising from highly excited \feii\ levels (from 30\,000 to 40\,000\,$\si{K}$; see Table~\ref{list lines faint} and Fig.~\ref{Fe II lines faint}) at wavelengths between 2400 and 2900 \A. A key feature of these lines is that their lower levels require much higher electronic densities to be populated at LTE, around $n_e$\,$\ga$\,$10^8 \, \si{cm^{-3}}$ \citep[ChiantiPy][]{ChiantiPy_2012}; at low densities, they tend to be less populated than at LTE. Thus, this set of lines can be used as a probe of the electronic density within the transiting cometary tails.
To do so, we started by measuring the mean absorption depths of the December 6, 1997 comet in the [+25; +40] km/s range for 26 high-energy \feii\ lines. Then, given the cometary cloud properties measured in Sect.~\ref{Temperature measurement} ($T$\,$\sim$\,$10\,500\,\si{K}$, $\overline{\alpha}$\,$\sim$\,$0.356$, $\gamma$\,$\sim$\,$4.72$), we used these absorption measurements to extract the relative abundances of the energy levels corresponding to those lines (Eq.~\eqref{equation diagram excitation}). These abundance estimates rely on the fact that the second equality in Eq.~\eqref{equation diagram excitation} remains valid for energy levels that are not populated at LTE, as long as \emph{\emph{most}} of the absorbing gas follows a Boltzmann distribution:
$$\frac{\text{N}_l}{g_l \text{N}_{\text{\feii}}}  =   \frac{- \ln(1-{\rm \overline{abs}}_{lu} \big/ \overline{\alpha})}{\gamma \frac{\lambda_{lu}}{\lambda_0} g_l f_{lu} Z(T)}.
$$

\begin{figure}[h]
    \centering    
    \includegraphics[clip, trim = 2 20 0 50 , scale = 0.4]{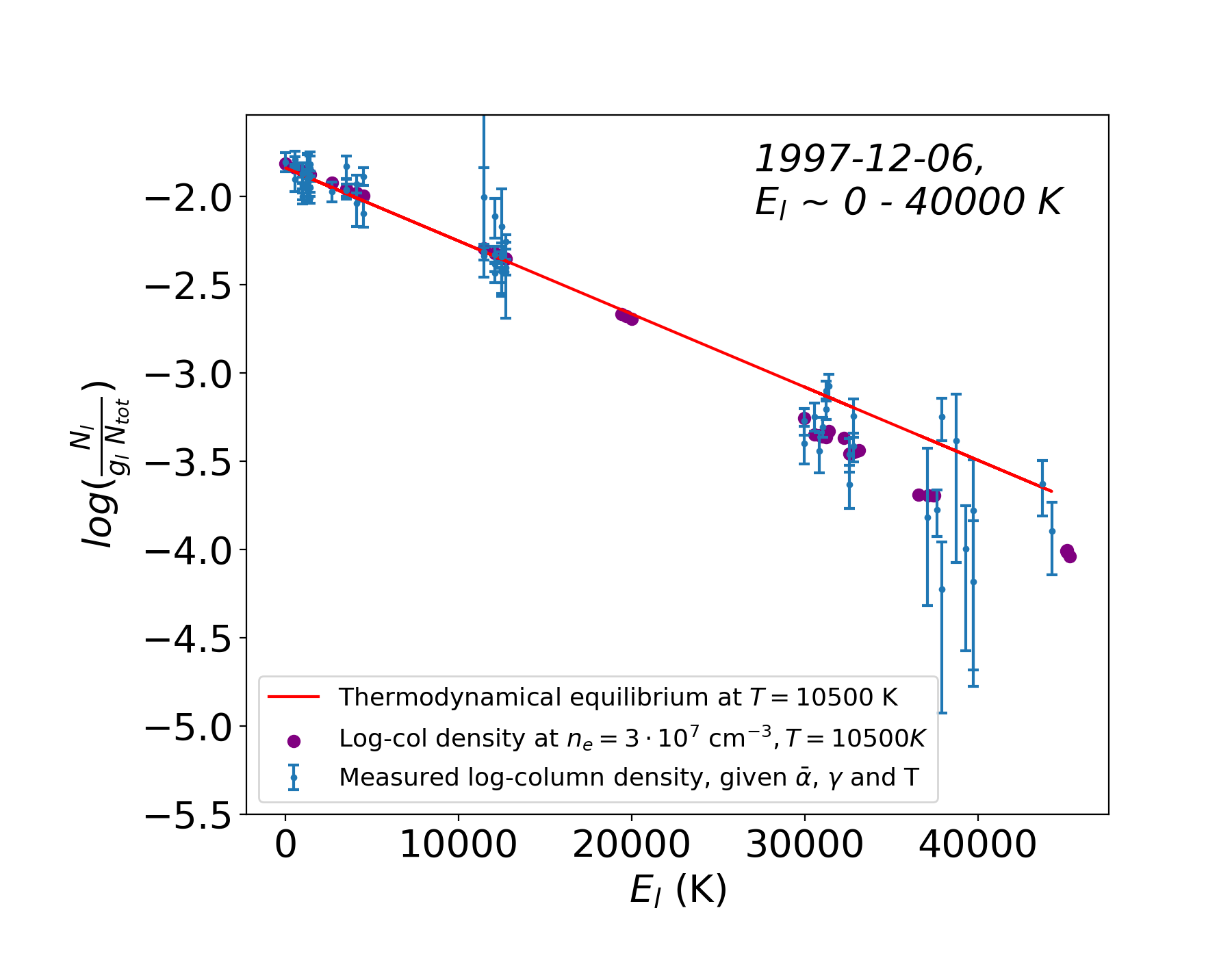}

    \caption[]{Full excitation diagram of \feii\ within December 6, 1997 comet showing measured column-density of various energy levels inferred from the measured absorptions and the parameters obtained in Sect.~\ref{Temperature measurement} (blue dots), the theoretical energy distribution at LTE (red line), and the theoretical energy distribution with an electron density of $3 \times 10^7$\,$\si{cm}^{-3}$ for a few representative energy levels as computed using ChiantiPy (purple dots).}
    \label{excitation diagram all lines}
\end{figure}

Then, we compared these relative abundances to the values predicted by LTE at the measured temperature ($10\,500 \ \si{K}$) by building the complete excitation diagram of \feii\ within the cometary tail (Fig.~\ref{excitation diagram all lines}). This diagram clearly shows that high-energy ($E_l$\,$\ga$\,$30\,000\,\si{K}$) \feii\ levels are under-populated compared to LTE, hinting that the electronic density is not high enough to impose a Boltzmann distribution.

Finally, in order to quantify the typical electronic density within the cometary cloud,
we compared the energy distribution of \feii\ measured in our spectra to the theoretical energy distribution for various electronic densities, obtained with ChiantiPy (to compute the distribution, a temperature of $10\,500 \ \si{K}$ was used, as measured in Sect. \ref{Temperature measurement}). Although it is difficult to give a precise estimate, the excitation diagram of \feii\ given in Fig.~\ref{excitation diagram all lines} seems to indicate an electronic density in the [2-4]$\times10^7 \, \si{cm^{-3}}$ range. These results are consistent with the work of \cite{Mouillet_1995}, where a minimum electronic density of $10^6 \ \si{cm^{-3}}$ in a transiting comet was inferred from its absorption in the \caii\ triplet.

\section{Discussion}
\label{Discussion}

\subsection{Deviation from the curve of growth model}

\begin{figure}[h!]
    \centering    
    \includegraphics[clip,  trim = 20 20 0 60, scale = 0.36]{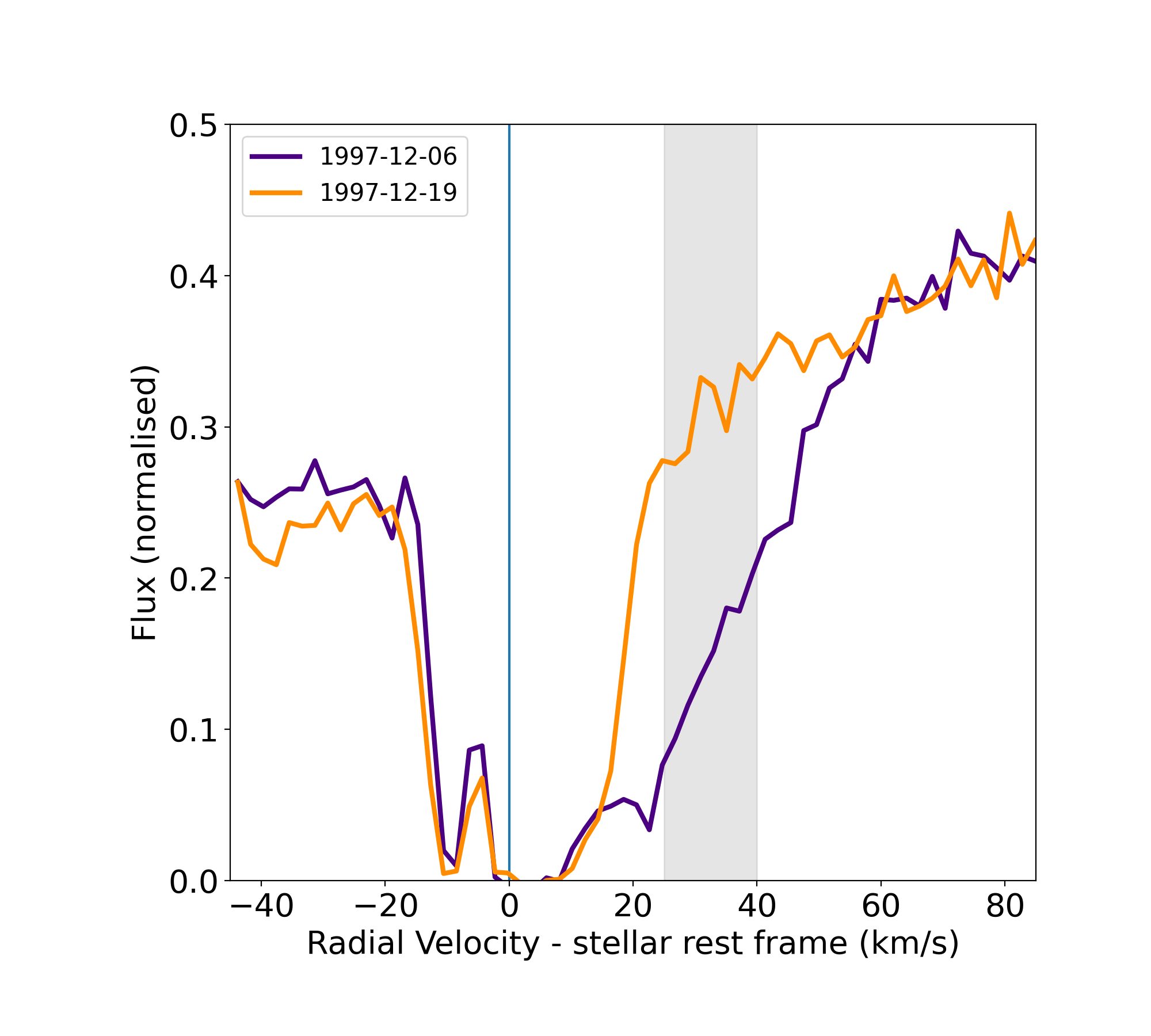}

    \caption[]{\feii\ line at 2600.17\,\A, showing a very deep cometary absorption between +25 and +40 km/s in the December 6, 1997 observations. The average absorption depth is about 55~\%, which is well above the size of the comet estimated using fainter lines from higher energy levels ($\sim 36 \%$). The deep absorption seen at -10~km/s relative to the \bp\ radial velocity is due to the interstellar medium, which is not seen in the other \feii\ lines arising from excited levels. }
    \label{2600_spectra}
\end{figure}

The analysis of the exocomet transit signature in the December 6, 1997 observation shows that the transiting gas in the [+12; +40] km/s range is described well by an homogeneous cloud covering, on average, 44\% of the stellar surface, with a total \feii\ column density of $\sim 1.1 \cdot 10^{15} \, \si{cm^{-2}}$, a temperature of $\sim$\,$10\,500 \, \si{K}$ (Sect.~\ref{Temperature measurement}), and an electronic density of approximately $10^7 \, \si{cm^{-3}}$ (Sect.~\ref{The electronic density}). These results were obtained by using most of the well-detected \feii\ lines in the 2300 - 2800 \A\ wavelength domain. However, we note that for a few very strong lines ($g_l f_{lu}$\,$>$\,$0.7$) rising from low energy levels ($E_l \leq 4500 \ \si{K}$), the measured absorption depth seems to deviate from the curve of growth model. For instance, the mean absorption depth of the December 6, 1997 comet in the [+25; +40]\,km/s range exceeds 50\% for the lines at 2383 \A\ ($g_l f_{lu} = 3.20$), 2396 \A\ ($g_l f_{lu} = 2.23$) and 2600 \A\ ($g_l f_{lu} = 2.63$, see Fig. \ref{2600_spectra}), which is well above the measured size of the cometary cloud ($\sim 36 \%$) in this radial velocity domain.

\begin{figure}[t]
    \centering    
    \includegraphics[clip,  trim = 30 0 0 50, scale = 0.44]{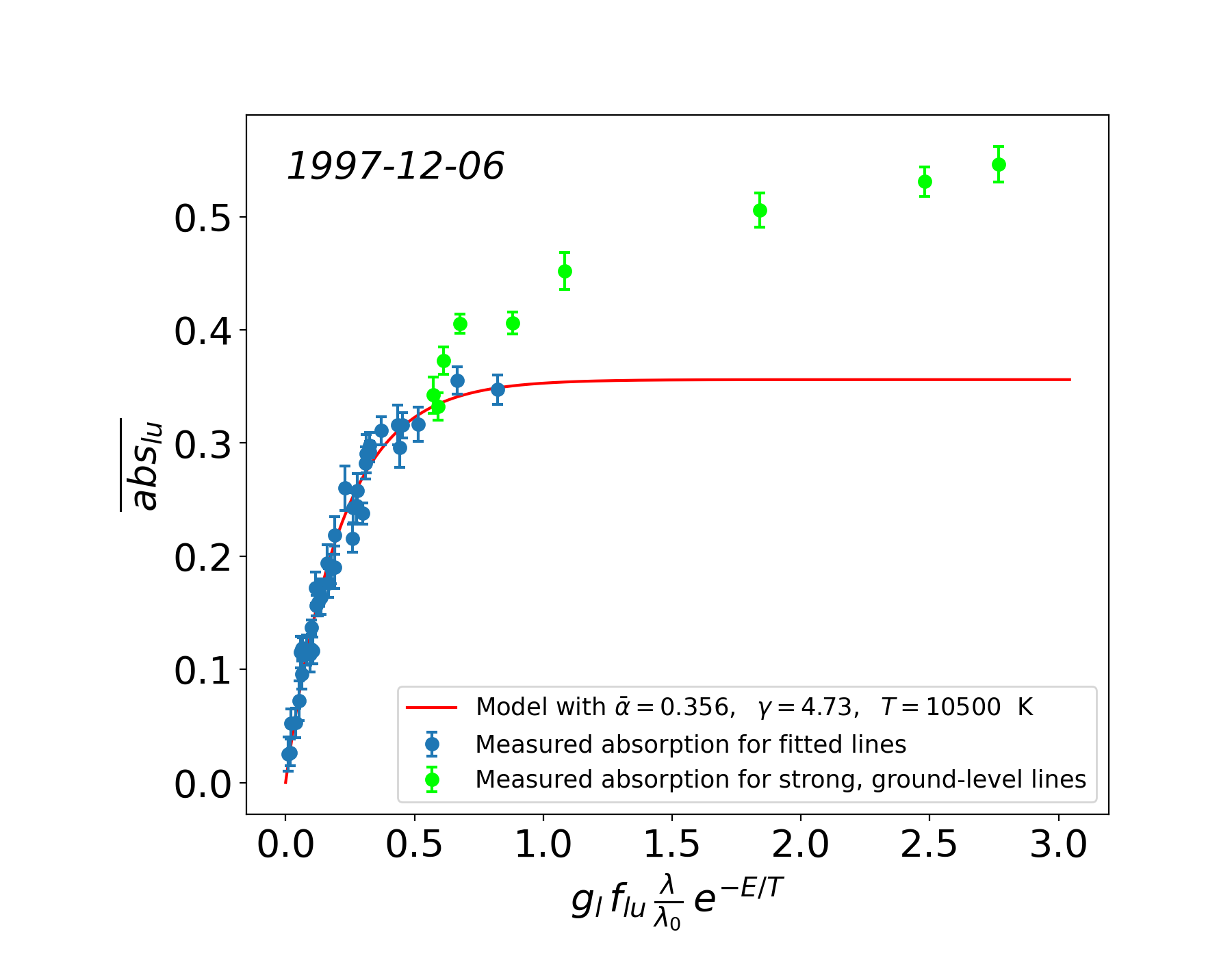}

    \caption[]{Comparison between the curve of growth model fitted in Sect.~\ref{Temperature and electron density} to the absorption of December 6, 1997 comet in weak \feii\ lines, and the measured absorption depths of this comet in the strongest ground-level \feii\ lines (green).}
    \label{o4g_all_lines}
\end{figure}

This discrepancy between the curve of growth model fitted in Sect.~\ref{Temperature measurement}, which matches the measured absorption depths on the "weak" \feii\ lines ($0.08 < g_l f_ul < 0.7$) strikingly well, and the absorption measured in the strongest ones ($g_l f_ul > 0.7$) can be revealed by plotting the curve of growth model with all the absorption measurements (Fig.~\ref{o4g_all_lines}). We see that while the cometary absorption depths in the weakest \feii\ lines (blue dots) saturate towards $36 \%$, the absorption in strong, ground-level \feii\ lines (green dots) keeps increasing slowly as the line strength ($\simeq g_l f_{lu}$) increases. This shows that the complexity of the transiting object is probably underestimated by our model. For instance, the absorption measurements might be better explained by the transit of at least two gaseous components: a dense one, which quickly saturates towards a 36 \% absorption depth as the line strength increases, and a thinner one, saturating more slowly, explaining why the absorption increases again by about $20 \%$ when we reach the strongest lines.

Although our simple model allows for a fairly accurate characterisation of transiting comets, it thus fails to probe the complex structure that such objects probably have. Nevertheless, our study shows the tremendous potential of the many near ultraviolet \feii\ lines for investigating the properties of the transiting gaseous tails. This paves the way for the development of more sophisticated models, in order to interpret the variable features observed in the spectrum of several stars in even greater detail. Such models may allow us to probe the temperature and density distribution of the transiting clouds - which are probably not perfectly homogeneous, as revealed by the study of the December 6, 1997 comet - as well as the orbital dynamics of the evaporating nucleus. In particular, one key objective of forthcoming works will be to infer the radial velocity of the nucleus from the absorption profile of its tail.

These models will also have to take into account the inhomogeneous distribution of the stellar flux at a given wavelength on the stellar disc. Indeed, this distribution is affected by limb darkening, which reduced the luminosity of the edges of the stellar disc, and by the rapid rotation velocity of stars such as \bp\ ($\sim$130\,$\si{km s^{-1}}$), which shifts the local stellar lines. Strictly speaking, occulting a fraction $\alpha$ of the stellar disc is thus not equivalent to reducing the stellar flux by a fraction $\alpha$  at each wavelength. This subtlety, which was neglected in our study, might actually be useful to constrain the 2D distribution of transiting cometary tails, by probing the location of the stellar regions they occult.

\subsection{Optical thickness profile}

The curve of growth model developed in Sect.~\ref{First Model} to derive the physical properties of \bp\ exocomets uses the hypothesis that the optical thickness - characterised by the $\beta$ parameter - is constant over the radial velocity range where the cometary absorption depths are measured. In fact, it can be shown that, even if the optical thickness profile is not flat, the model still provides very good estimates of the size, temperature, and column density of the absorbing cloud. To check this, we generated artificial absorption spectra of transiting comets with known column densities, and with quickly varying optical thickness over radial velocity. We found that, as long as the optical thickness stays below unity, or does not vary by a large factor ($\leq 10$) over the studied radial velocity range, the curve of growth model provides very good estimates of the true physical parameters of the absorbing gas ($N_{tot}, \alpha, T$).

To go further, we could also subdivide the radial velocity domain [$v_1$, $v_2$] into elementary intervals and apply the curve of growth model to the absorption depths measured within every range [$v$, $v + dv$]. This would allow us to probe the radial velocity dependency of the size ($\alpha(v)$) and optical thickness ($\gamma(v)$) of the gaseous cometary cloud. Such an analysis could thus give a first insight into the geometry of a transiting gaseous tail, probing, in particular, how its size and optical thickness evolve as the gas is pushed away from the star by the radiation pressure. However, this analysis is far beyond the scope of the present work.

\section{Conclusion}
\label{Conclusion}

Using HST archival UV spectra of \bp, we detected variable absorptions in dozens of \feii\ lines, which are due to the transit of gaseous cometary tails. To better characterise the properties of these objects, we developed a simple model that describes the absorbing gas with a homogeneous cloud, covering a fraction $\alpha$ of the stellar disc. This model matches the measured absorptions in most of the \feii\ lines remarkably well - with the exception of the strongest ones - showing that the variable features seen in the \bp \ spectrum are well interpreted by the transit of large ($\overline{\alpha} \sim 0.4$), dense (N$_{\text{\feii}} \sim 10^{15} \si{cm^{-2}}$) and hot ($T \geq 5000 \ \si{K}$) cometary gas tails. In the comet that transited in December 6, 1997, the population of the energy levels is consistent with LTE, which allows an estimate of the temperature within the gas tail of $10\,500 \pm 500 \, \si{K}$. Using \feii\ lines arising from highly excited electronic levels, we estimated the typical electronic density to be around a few $10^7 \ \si{cm^{-3}}$ in this cometary gas cloud. All the results obtained here are consistent with previous observational studies, notably with the quantitative estimates given by \cite{Lagrange_1995}, \cite{Mouillet_1995}, and \cite{Kiefer_2014}. They are, however, slightly different to the theoretical predictions of \cite{Beust_1993}, for which the formation of a shock front with very high temperature ($T \sim 10^{5} \ \si{K}$) and rather low density ($n_e \sim 10^{3} - 10^{6}$) is necessary to explain the presence of highly ionised species (such as \aliii) in the transiting cometary tails. Further work will be needed to better understand this discrepancy. 

Our results also confirm the huge difference between \bp\ and solar comets; while the iron in \bp\ comets is mainly present as Fe$^+$ (only very shallow cometary absorption is detected in \fei\ lines; see \cite{Welsh2016})), with column densities of the order of $10^{15} \si{cm^{-2}}$, solar comets contain mainly neutral iron, with much lower column densities ($\sim 10^8 - 10^{10} \ \si{cm^{-2}}$, \cite{Manfroid2021}). This is likely to be a consequence of the very short distance between \bp\ transiting comets and their star, which induces an intense sublimation of metal-rich material at the surface of the nuclei, while the tails of solar comets mainly result from the sublimation of volatile ices.

Finally, the curve of growth model appears to be an efficient tool to characterise the gaseous clouds surrounding transiting exocomets, detected through variable spectroscopic absorptions. The exocomet curve of growth is the analogue of the classical curve of growth used for interstellar clouds and stellar atmospheres. A fit to this curve provides quantitative estimates of the typical size of the absorbing cloud, and the column densities of the detected species. When several lines from various excitation energy levels are available, the temperature and possibly the electronic density can also be constrained. Further works using this new tool are already in progress.  

\begin{acknowledgements}
T.V., A.L., G.H. \& A.V.-M acknowledge funding from the Centre National d’\'Etudes Spatiales (CNES). 
\end{acknowledgements}

\bibliographystyle{aa}
\bibliography{main}

\begin{thebibliography}{42}
\expandafter\ifx\csname natexlab\endcsname\relax\def\natexlab#1{#1}\fi

\bibitem[{{Apai} {et~al.}(2015){Apai}, {Schneider}, {Grady}, {Wyatt}, {Lagrange}, {Kuchner}, {Stark}, \& {Lubow}}]{Apai_2015}
{Apai}, D., {Schneider}, G., {Grady}, C.~A., {et~al.} 2015, \apj, 800, 136

\bibitem[{{Beust} {et~al.}(1990){Beust}, {Lagrange-Henri}, {Vidal-Madjar}, \& {Ferlet}}]{Beust_1990}
{Beust}, H., {Lagrange-Henri}, A.~M., {Vidal-Madjar}, A., \& {Ferlet}, R. 1990, \aap, 236, 202

\bibitem[{{Beust} \& {Tagger}(1993)}]{Beust_1993}
{Beust}, H. \& {Tagger}, M. 1993, \icarus, 106, 42

\bibitem[{{Brandeker}(2011)}]{Brandeker2011}
{Brandeker}, A. 2011, \apj, 729, 122

\bibitem[{{Brandeker} {et~al.}(2016){Brandeker}, {Cataldi}, {Olofsson}, {Vandenbussche}, {Acke}, {Barlow}, {Blommaert}, {Cohen}, {Dent}, {Dominik}, {Di Francesco}, {Fridlund}, {Gear}, {Glauser}, {Greaves}, {Harvey}, {Heras}, {Hogerheijde}, {Holland}, {Huygen}, {Ivison}, {Leeks}, {Lim}, {Liseau}, {Matthews}, {Pantin}, {Pilbratt}, {Royer}, {Sibthorpe}, {Waelkens}, \& {Walker}}]{Brandeker_2016}
{Brandeker}, A., {Cataldi}, G., {Olofsson}, G., {et~al.} 2016, \aap, 591, A27

\bibitem[{{Ferlet} {et~al.}(1987){Ferlet}, {Hobbs}, \& {Vidal-Madjar}}]{Ferlet_1987}
{Ferlet}, R., {Hobbs}, L.~M., \& {Vidal-Madjar}, A. 1987, \aap, 185, 267

\bibitem[{{Gontcharov}(2007)}]{Gontcharov2007}
{Gontcharov}, G.~A. 2007, VizieR Online Data Catalog, III/252

\bibitem[{{Jolly} {et~al.}(1998){Jolly}, {McPhate}, {Lecavelier}, {Lagrange}, {Lemaire}, {Feldman}, {Vidal Madjar}, {Ferlet}, {Malmasson}, \& {Rostas}}]{Jolly1998}
{Jolly}, A., {McPhate}, J.~B., {Lecavelier}, A., {et~al.} 1998, \aap, 329, 1028

\bibitem[{{Kalas} {et~al.}(2000){Kalas}, {Larwood}, {Smith}, \& {Schultz}}]{Kalas_2001}
{Kalas}, P., {Larwood}, J., {Smith}, B.~A., \& {Schultz}, A. 2000, \apjl, 530, L133

\bibitem[{{Kennedy}(2018)}]{Kennedy_2018}
{Kennedy}, G.~M. 2018, \mnras, 479, 1997

\bibitem[{{Kiefer} {et~al.}(2014{\natexlab{a}}){Kiefer}, {Lecavelier des Etangs}, {Augereau}, {Vidal-Madjar}, {Lagrange}, \& {Beust}}]{Kiefer_2014b}
{Kiefer}, F., {Lecavelier des Etangs}, A., {Augereau}, J.~C., {et~al.} 2014{\natexlab{a}}, \aap, 561, L10

\bibitem[{{Kiefer} {et~al.}(2014{\natexlab{b}}){Kiefer}, {Lecavelier des Etangs}, {Boissier}, {Vidal-Madjar}, {Beust}, {Lagrange}, {H{\'e}brard}, \& {Ferlet}}]{Kiefer_2014}
{Kiefer}, F., {Lecavelier des Etangs}, A., {Boissier}, J., {et~al.} 2014{\natexlab{b}}, \nat, 514, 462

\bibitem[{{Kiefer} {et~al.}(2023){Kiefer}, {Van Grootel}, {Lecavelier des Etangs}, {Szab{\'o}}, {Brandeker}, {Broeg}, {Collier Cameron}, {Deline}, {Olofsson}, {Wilson}, {Sousa}, {Gandolfi}, {H{\'e}brard}, {Alibert}, {Alonso}, {Anglada}, {B{\'a}rczy}, {Barrado}, {Barros}, {Baumjohann}, {Beck}, {Beck}, {Benz}, {Billot}, {Bonfils}, {Cabrera}, {Charnoz}, {Csizmadia}, {Davies}, {Deleuil}, {Delrez}, {Demangeon}, {Demory}, {Ehrenreich}, {Erikson}, {Fortier}, {Fossati}, {Fridlund}, {Gillon}, {G{\"u}del}, {Heng}, {Hoyer}, {Isaak}, {Kiss}, {Laskar}, {Lendl}, {Lovis}, {Magrin}, {Maxted}, {Munari}, {Nascimbeni}, {Ottensamer}, {Pagano}, {Pall{\'e}}, {Peter}, {Piazza}, {Piotto}, {Pollacco}, {Queloz}, {Ragazzoni}, {Rando}, {Ratti}, {Rauer}, {Reimers}, {Ribas}, {Santos}, {Scandariato}, {S{\'e}gransan}, {Simon}, {Smith}, {Steller}, {Thomas}, {Udry}, {Walter}, \& {Walton}}]{Kiefer_2023}
{Kiefer}, F., {Van Grootel}, V., {Lecavelier des Etangs}, A., {et~al.} 2023, \aap, 671, A25

\bibitem[{{Kiefer} {et~al.}(2019){Kiefer}, {Vidal-Madjar}, {Lecavelier des Etangs}, {Bourrier}, {Ehrenreich}, {Ferlet}, {H{\'e}brard}, \& {Wilson}}]{Kiefer2019}
{Kiefer}, F., {Vidal-Madjar}, A., {Lecavelier des Etangs}, A., {et~al.} 2019, \aap, 621, A58

\bibitem[{Kramida {et~al.}(2023)Kramida, {Yu.~Ralchenko}, Reader, \& {and NIST ASD Team}}]{NIST_ASD}
Kramida, A., {Yu.~Ralchenko}, Reader, J., \& {and NIST ASD Team}. 2023, {NIST Atomic Spectra Database (ver. 5.11), [Online]. Available: {\tt{https://physics.nist.gov/asd}} [2023, December 15]. National Institute of Standards and Technology, Gaithersburg, MD.}

\bibitem[{{Lacour} {et~al.}(2021){Lacour}, {Wang}, {Rodet}, {Nowak}, {Shangguan}, {Beust}, {Lagrange}, {Abuter}, {Amorim}, {Asensio-Torres}, {Benisty}, {Berger}, {Blunt}, {Boccaletti}, {Bohn}, {Bolzer}, {Bonnefoy}, {Bonnet}, {Bourdarot}, {Brandner}, {Cantalloube}, {Caselli}, {Charnay}, {Chauvin}, {Choquet}, {Christiaens}, {Cl{\'e}net}, {Coud{\'e} du Foresto}, {Cridland}, {Dembet}, {Dexter}, {de Zeeuw}, {Drescher}, {Duvert}, {Eckart}, {Eisenhauer}, {Gao}, {Garcia}, {Garcia Lopez}, {Gendron}, {Genzel}, {Gillessen}, {Girard}, {Haubois}, {Hei{\ss}el}, {Henning}, {Hinkley}, {Hippler}, {Horrobin}, {Houll{\'e}}, {Hubert}, {Jocou}, {Kammerer}, {Keppler}, {Kervella}, {Kreidberg}, {Lapeyr{\`e}re}, {Le Bouquin}, {L{\'e}na}, {Lutz}, {Maire}, {M{\'e}rand}, {Molli{\`e}re}, {Monnier}, {Mouillet}, {Nasedkin}, {Ott}, {Otten}, {Paladini}, {Paumard}, {Perraut}, {Perrin}, {Pfuhl}, {Rickman}, {Pueyo}, {Rameau}, {Rousset}, {Rustamkulov}, {Samland}, {Shimizu}, {Sing}, {Stadler}, {Stolker}, {Straub}, {Straubmeier}, {Sturm},
  {Tacconi}, {van Dishoeck}, {Vigan}, {Vincent}, {von Fellenberg}, {Ward-Duong}, {Widmann}, {Wieprecht}, {Wiezorrek}, {Woillez}, {Yazici}, {Young}, \& {the GRAVITY Collaboration}}]{Lacour_2021}
{Lacour}, S., {Wang}, J.~J., {Rodet}, L., {et~al.} 2021, arXiv e-prints, arXiv:2109.10671

\bibitem[{{Lagrange} {et~al.}(2010){Lagrange}, {Bonnefoy}, {Chauvin}, {Apai}, {Ehrenreich}, {Boccaletti}, {Gratadour}, {Rouan}, {Mouillet}, {Lacour}, \& {Kasper}}]{Lagrange_2010}
{Lagrange}, A.~M., {Bonnefoy}, M., {Chauvin}, G., {et~al.} 2010, Science, 329, 57

\bibitem[{{Lagrange} {et~al.}(2019){Lagrange}, {Meunier}, {Rubini}, {Keppler}, {Galland}, {Chapellier}, {Michel}, {Balona}, {Beust}, {Guillot}, {Grandjean}, {Borgniet}, {M{\'e}karnia}, {Wilson}, {Kiefer}, {Bonnefoy}, {Lillo-Box}, {Pantoja}, {Jones}, {Iglesias}, {Rodet}, {Diaz}, {Zapata}, {Abe}, \& {Schmider}}]{Lagrange_2019}
{Lagrange}, A.~M., {Meunier}, N., {Rubini}, P., {et~al.} 2019, Nature Astronomy, 3, 1135

\bibitem[{{Lagrange} {et~al.}(1995){Lagrange}, {Vidal-Madjar}, {Deleuil}, {Emerich}, {Beust}, \& {Ferlet}}]{Lagrange_1995}
{Lagrange}, A.~M., {Vidal-Madjar}, A., {Deleuil}, M., {et~al.} 1995, \aap, 296, 499

\bibitem[{{Lagrange-Henri} {et~al.}(1989){Lagrange-Henri}, {Beust}, {Ferlet}, \& {Vidal-Madjar}}]{Lagrange_1989}
{Lagrange-Henri}, A.~M., {Beust}, H., {Ferlet}, R., \& {Vidal-Madjar}, A. 1989, \aap, 215, L5

\bibitem[{{Landi} {et~al.}(2012){Landi}, {Del Zanna}, {Young}, {Dere}, \& {Mason}}]{ChiantiPy_2012}
{Landi}, E., {Del Zanna}, G., {Young}, P.~R., {Dere}, K.~P., \& {Mason}, H.~E. 2012, \apj, 744, 99

\bibitem[{{Lecavelier des Etangs} {et~al.}(2022){Lecavelier des Etangs}, {Cros}, {H{\'e}brard}, {Martioli}, {Duquesnoy}, {Kenworthy}, {Kiefer}, {Lacour}, {Lagrange}, {Meunier}, \& {Vidal-Madjar}}]{Lecavelier_2022}
{Lecavelier des Etangs}, A., {Cros}, L., {H{\'e}brard}, G., {et~al.} 2022, Scientific Reports, 12, 5855

\bibitem[{{Lecavelier des Etangs} {et~al.}(2001){Lecavelier des Etangs}, {Vidal-Madjar}, {Roberge}, {Feldman}, {Deleuil}, {Andr{\'e}}, {Blair}, {Bouret}, {D{\'e}sert}, {Ferlet}, {Friedman}, {H{\'e}brard}, {Lemoine}, \& {Moos}}]{Lecavelier2001}
{Lecavelier des Etangs}, A., {Vidal-Madjar}, A., {Roberge}, A., {et~al.} 2001, \nat, 412, 706

\bibitem[{{Manfroid} {et~al.}(2021){Manfroid}, {Hutsem{\'e}kers}, \& {Jehin}}]{Manfroid2021}
{Manfroid}, J., {Hutsem{\'e}kers}, D., \& {Jehin}, E. 2021, \nat, 593, 372

\bibitem[{{Miles} {et~al.}(2016){Miles}, {Roberge}, \& {Welsh}}]{Miles_2016}
{Miles}, B.~E., {Roberge}, A., \& {Welsh}, B. 2016, \apj, 824, 126

\bibitem[{{Miret-Roig} {et~al.}(2020){Miret-Roig}, {Galli}, {Brandner}, {Bouy}, {Barrado}, {Olivares}, {Antoja}, {Romero-G{\'o}mez}, {Figueras}, \& {Lillo-Box}}]{Miret-Roig_2020}
{Miret-Roig}, N., {Galli}, P.~A.~B., {Brandner}, W., {et~al.} 2020, \aap, 642, A179

\bibitem[{{Montgomery} \& {Welsh}(2012)}]{Montgomery2012}
{Montgomery}, S.~L. \& {Welsh}, B.~Y. 2012, \pasp, 124, 1042

\bibitem[{{Mouillet} \& {Lagrange}(1995)}]{Mouillet_1995}
{Mouillet}, D. \& {Lagrange}, A.~M. 1995, \aap, 297, 175

\bibitem[{{Nowak} {et~al.}(2020){Nowak}, {Lacour}, {Lagrange}, {Rubini}, {Wang}, {Stolker}, {Abuter}, {Amorim}, {Asensio-Torres}, {Baub{\"o}ck}, {Benisty}, {Berger}, {Beust}, {Blunt}, {Boccaletti}, {Bonnefoy}, {Bonnet}, {Brandner}, {Cantalloube}, {Charnay}, {Choquet}, {Christiaens}, {Cl{\'e}net}, {Coud{\'e} Du Foresto}, {Cridland}, {de Zeeuw}, {Dembet}, {Dexter}, {Drescher}, {Duvert}, {Eckart}, {Eisenhauer}, {Gao}, {Garcia}, {Garcia Lopez}, {Gardner}, {Gendron}, {Genzel}, {Gillessen}, {Girard}, {Grandjean}, {Haubois}, {Hei{\ss}el}, {Henning}, {Hinkley}, {Hippler}, {Horrobin}, {Houll{\'e}}, {Hubert}, {Jim{\'e}nez-Rosales}, {Jocou}, {Kammerer}, {Kervella}, {Keppler}, {Kreidberg}, {Kulikauskas}, {Lapeyr{\`e}re}, {Le Bouquin}, {L{\'e}na}, {M{\'e}rand}, {Maire}, {Molli{\`e}re}, {Monnier}, {Mouillet}, {M{\"u}ller}, {Nasedkin}, {Ott}, {Otten}, {Paumard}, {Paladini}, {Perraut}, {Perrin}, {Pueyo}, {Pfuhl}, {Rameau}, {Rodet}, {Rodr{\'\i}guez-Coira}, {Rousset}, {Scheithauer}, {Shangguan}, {Stadler}, {Straub},
  {Straubmeier}, {Sturm}, {Tacconi}, {van Dishoeck}, {Vigan}, {Vincent}, {von Fellenberg}, {Ward-Duong}, {Widmann}, {Wieprecht}, {Wiezorrek}, {Woillez}, \& {GRAVITY Collaboration}}]{Nowak2020}
{Nowak}, M., {Lacour}, S., {Lagrange}, A.~M., {et~al.} 2020, \aap, 642, L2

\bibitem[{{Pavlenko} {et~al.}(2022){Pavlenko}, {Kulyk}, {Shubina}, {Vasylenko}, {Dobrycheva}, \& {Korsun}}]{Pavlenko_2022}
{Pavlenko}, Y., {Kulyk}, I., {Shubina}, O., {et~al.} 2022, \aap, 660, A49

\bibitem[{{Rebollido} {et~al.}(2020){Rebollido}, {Eiroa}, {Montesinos}, {Maldonado}, {Villaver}, {Absil}, {Bayo}, {Canovas}, {Carmona}, {Chen}, {Ertel}, {Henning}, {Iglesias}, {Launhardt}, {Liseau}, {Meeus}, {Mo{\'o}r}, {Mora}, {Olofsson}, {Rauw}, \& {Riviere-Marichalar}}]{Rebollido2020}
{Rebollido}, I., {Eiroa}, C., {Montesinos}, B., {et~al.} 2020, \aap, 639, A11

\bibitem[{{Roberge} {et~al.}(2000){Roberge}, {Feldman}, {Lagrange}, {Vidal-Madjar}, {Ferlet}, {Jolly}, {Lemaire}, \& {Rostas}}]{Roberge2000}
{Roberge}, A., {Feldman}, P.~D., {Lagrange}, A.~M., {et~al.} 2000, \apj, 538, 904

\bibitem[{{Roberge} {et~al.}(2006){Roberge}, {Feldman}, {Weinberger}, {Deleuil}, \& {Bouret}}]{Roberge_2006}
{Roberge}, A., {Feldman}, P.~D., {Weinberger}, A.~J., {Deleuil}, M., \& {Bouret}, J.-C. 2006, \nat, 441, 724

\bibitem[{{Roberge} {et~al.}(2014){Roberge}, {Welsh}, {Kamp}, {Weinberger}, \& {Grady}}]{Roberge_2014}
{Roberge}, A., {Welsh}, B.~Y., {Kamp}, I., {Weinberger}, A.~J., \& {Grady}, C.~A. 2014, \apjl, 796, L11

\bibitem[{{Smith} \& {Terrile}(1984)}]{Smith_1984}
{Smith}, B.~A. \& {Terrile}, R.~J. 1984, Science, 226, 1421

\bibitem[{{Snellen} \& {Brown}(2018)}]{Snellen_2018}
{Snellen}, I.~A.~G. \& {Brown}, A.~G.~A. 2018, Nature Astronomy, 2, 883

\bibitem[{{Str{\o}m} {et~al.}(2020){Str{\o}m}, {Bodewits}, {Knight}, {Kiefer}, {Jones}, {Kral}, {Matr{\`a}}, {Bodman}, {Capria}, {Cleeves}, {Fitzsimmons}, {Haghighipour}, {Harrison}, {Iglesias}, {Kama}, {Linnartz}, {Majumdar}, {de Mooij}, {Milam}, {Opitom}, {Rebollido}, {Rogers}, {Snodgrass}, {Sousa-Silva}, {Xu}, {Lin}, \& {Zieba}}]{Strom_2020}
{Str{\o}m}, P.~A., {Bodewits}, D., {Knight}, M.~M., {et~al.} 2020, \pasp, 132, 101001

\bibitem[{{Vidal-Madjar} {et~al.}(1986){Vidal-Madjar}, {Hobbs}, {Ferlet}, {Gry}, \& {Albert}}]{Vidal-Majar_1986}
{Vidal-Madjar}, A., {Hobbs}, L.~M., {Ferlet}, R., {Gry}, C., \& {Albert}, C.~E. 1986, \aap, 167, 325

\bibitem[{{Vidal-Madjar} {et~al.}(1994){Vidal-Madjar}, {Lagrange-Henri}, {Feldman}, {Beust}, {Lissauer}, {Deleuil}, {Ferlet}, {Gry}, {Hobbs}, {McGrath}, {McPhate}, \& {Moos}}]{Vidal-Madjar_1994}
{Vidal-Madjar}, A., {Lagrange-Henri}, A.~M., {Feldman}, P.~D., {et~al.} 1994, \aap, 290, 245

\bibitem[{{Welsh} \& {Montgomery}(2013)}]{Welsh2013}
{Welsh}, B.~Y. \& {Montgomery}, S. 2013, \pasp, 125, 759

\bibitem[{{Welsh} \& {Montgomery}(2016)}]{Welsh2016}
{Welsh}, B.~Y. \& {Montgomery}, S. 2016, \pasp, 128, 064201

\bibitem[{{Zieba} {et~al.}(2019){Zieba}, {Zwintz}, {Kenworthy}, \& {Kennedy}}]{Zieba_2019}
{Zieba}, S., {Zwintz}, K., {Kenworthy}, M.~A., \& {Kennedy}, G.~M. 2019, \aap, 625, L13

\end{thebibliography}

\begin{appendix}
\section{Line and energy level parameters}

\vspace{0.5 cm}

\begin{table}[h!]
    \centering 
    \begin{threeparttable}

    \caption{Parameters of the strongest \feii\ lines in the three main series at 2400,  2600, and 2750\,\A\ (see Sect. \ref{First Model} for details).}

    \begin{tabular}{ r  r  r  c  c }          
            \hline                     

            $\lambda_{lu}$  & $E_l$  & $E_l/k_B$ & $A_{ul}$ \tnote{b}  & $g_l \ f_{lu}$ \tnote{b} \\
            (\A) & ($\si{cm^{-1}}) $  & ($\si{K}$) & ($10^7 \ \si{s^{-1}}$) &  \\

            \hline
            2328.11 & 668 & 960 & 6.60 & 0.215 \\
            2332.02     & 1873 & 2694 & 3.17 & 0.207 \\
            2333.51     & 385 & 554 & 13.1 & 0.643 \\
            2338.73     & 863 & 1242 & 11.3 & 0.371 \\
            2345.00 & 977 & 1406 & 9.27 & 0.306 \\
            2349.02 & 668 & 961  & 11.5 & 0.572 \\
            2355.61 & 2838 & 4084 & 2.67 & 0.089 \\
            2359.83     & 863 & 1242 & 5.00 & 0.251 \\
            2361.02 & 2430 & 3497 & 6.20 & 0.311 \\
            2362.74 & 2430  & 3497 & 1.41 & 0.095 \\
            2365.55     & 385 & 554 & 5.90 & 0.387\\
            2369.32     & 2838 & 4084 & 6.06 & 0.204\\
            2374.46     & 0 & 0 & 4.25 & 0.360 \\
            2375.92     & 3117 & 4486 & 9.80 & 0.166\\
            2380.00     & 2430 & 3497 &2.73 & 0.186 \\
            2381.49     & 668 & 961 & 3.10 & 0.211 \\
            \tnote{a} \ 2382.76    & 0 & 0 & 31.3 & 3.203 \\
            2385.11 & 3117 & 4486 & 3.20 & 0.109 \\
            \tnote{a} \ 2389.36    & 385 & 554 & 10.5 & 0.720 \\
            \tnote{a} \ 2396.35 & 385 & 554 & 25.9 & 2.234\\
            \tnote{a} \ 2399.97 & 668 & 961 & 13.9 & 0.722\\
            \tnote{a} \ 2405.62    & 668 & 961 & 19.6 & 1.363 \\
            2407.39     & 863 & 1242 & 16.1 & 0.561 \\
            2411,80     & 977 & 1406 & 23.7 & 0.414 \\
            2414.04 & 977 & 1406 & 10.2 & 0.357 \\

            \\

            2564.24     & 8392 & 12076 & 15.1 & 0.597 \\
            2567.68 & 8680 & 12492 & 11.0 & 0.218 \\
            2578.69     & 8847 & 12731 & 12.0 & 0.240 \\
            2583.36     & 8680 & 12492 & 8.80 & 0.353 \\
            \tnote{a} \ 2586.64 & 0 & 0 & 9.94 & 0.719 \\
            2592.31     & 8392 & 12076 & 5.70 & 0.345 \\
            \tnote{a} \ 2600.17    & 0 & 0 & 23.5 & 2.386 \\
            \tnote{a} \ 2607.87 & 668 & 961 & 17.3 & 0.707\\
            \tnote{a} \ 2612.65 & 385 & 554 & 12.0 & 0.984\\
            2614.60     & 863 & 1241 & 21.2 & 0.435 \\
            2618.40     & 668 & 961 & 4.88 & 0.302 \\
            2622.45     & 977 & 1406 & 5.60 & 0.116\\
            2626.45     & 385 & 554 & 3.52 & 0.365 \\
            2629.08     & 977 & 1406 & 8.74 & 0.363\\
            2632.11 & 668 & 961 & 6.29 & 0.524 \\

                \hline  
  
        \end{tabular} 
        
    \begin{tablenotes}
      \item[a] \small Lines discussed in Sect. 6 for the December 6, 1997 transiting comet.
      \item[b] \small The $A$ and $g \cdot f$ values of the lines were found on the NIST database (\cite{NIST_ASD}). The uncertainties are of the order of 10 \%.
    \end{tablenotes}
    
\label{list lines}

\end{threeparttable}
\end{table}

\newpage

\begin{table}[h!]
        \centering  
        \begin{threeparttable}

        \vspace{1 cm}

        \begin{tabular}{ r  r  r  c  c }          
            \hline                     

		  $\lambda_{lu}$ & $E_l$  & $E_l/k_B$ & $A_{ul}$  & $g_l \ f_{lu}$ \\
		  (\A) & ($\si{cm^{-1}}) $  & ($\si{K}$) & ($10^7 \ \si{s^{-1}}$) &  \\

            \hline

            2715.22 & 7955 & 11448	& 5.70 & 0.379	\\
            2725.69 & 8392 & 12077	& 0.96 & 0.064  \\
            2728.35	& 8392 & 12077	& 9.38 & 0.420	\\
            2731.54 & 8680 & 12492  & 2.79 & 0.125  \\
            2737.78	& 8680 & 12492 	& 12.2 & 0.275  \\
            2740.36	& 7955 & 11448  & 22.1 & 1.994  \\
            2744.01	& 8847 & 12732  & 19.7 & 0.891  \\
            2747.79	& 8392 & 12077  & 16.9 & 1.150  \\
            2756.55	& 7955 & 11448  & 21.5 & 2.454  \\ 
            2762.66	& 8847 & 12732  & 1.38 & 0.063  \\
            2769.75	& 8680 & 12492  & 0.48 & 0.033  \\
        \hline

        \end{tabular}   
        
    \begin{tablenotes}
      \small
      \centering
      \item Table \ref{list lines}. continued.
    \end{tablenotes}

    \label{list lines faint}

    \end{threeparttable}

\end{table}

\vspace{0.9 cm}

\begin{table}[h!]
        \centering  
        \begin{threeparttable}

        \caption{Same parameters for the faint \feii\ lines rising from highly excited levels, used in Sect.~\ref{The electronic density}.}

        \begin{tabular}{ r  r  r  c  c }          
            \hline                     

            $\lambda_{lu}$ & $E_l$  & $E_l/k_B$ & $A_{ul}$\tnote{a}  & $g_l \ f_{lu}$\tnote{a} \\
            (\A) & ($\si{cm^{-1}}) $  & ($\si{K}$) & ($10^7 \ \si{s^{-1}}$) &  \\

            \hline

            2424.88     & 22637 & 32578 & 22.1 & 2.34 \\
            2430.82     & 22810 & 32827 & 19.1 & 1.70 \\
            2445.26     & 20830 & 29977 & 27.8 & 2.00 \\
            2446.31 & 21812 & 31390 & 20.7 & 1.12 \\
            2471.42     & 22810 & 32827 & 15.4 & 0.85 \\
            2499.65     & 21581 & 31058 & 21.2 & 2.39 \\
            2512.52     & 21712 & 31246 & 23.0 & 2.18 \\
            2526.15     & 21251 & 30583 & 19.1 & 2.56 \\
            2527.05     & 20830 & 29977 & 24.7 & 1.42 \\
            2530.31     & 22637 & 32578 & 22.0 & 2.12 \\
            2530.32     & 21812 & 31390 & 20.2 & 0.78 \\
            2534.39     & 21430 & 30841 & 19.2 & 2.22 \\
            2535.18     & 21711 & 31245 & 18.3 & 1.41 \\

            \\
            
            2667.43     & 27620 & 39748 & 18.7 & 1.60 \\
            2685.55     & 30764 & 44273 & 15.7 & 1.70 \\
            2693.40     & 30388 & 43732 & 14.0 & 1.83 \\
            2704.79     & 27314 & 39308 & 13.8 & 1.21 \\
            2712.65     & 25428 & 36594 & 4.3 & 0.68 \\
            2717.02     & 27620 & 39748 & 11.5 & 0.77 \\
            2754.10     & 26352 & 37923 & 18.9 & 2.58 \\
            2768.32     & 26170 & 37662 & 15.8 & 2.55 \\
            2780.12     & 26352 & 37923 & 10.0 & 0.93 \\
            2784.51     & 26170 & 37662 & 10.6 & 1.23 \\
            2832.39     & 25787 & 37110 & 7.6 & 0.55 \\
            2841.48     & 26932 & 38758 & 7.6 & 0.37 \\
            2859.18     & 30388 & 43732 & 4.8 & 0.71 \\
        \hline

        \end{tabular}   
        
    \begin{tablenotes}
      \small
      \item[a] Parameters obtained from the NIST database. The uncertainties on $A_{ul}$ and $f_{lu}$ are typically of the order of 10 to 15 \%.
    \end{tablenotes}

    \label{list lines faint}

    \end{threeparttable}

\end{table}

\newpage

\begin{table}[h!]
        \centering 

        \begin{threeparttable}

        \caption{ First energy levels of \feii, which give rise to the three main line series at 2400, 2600, and 2750 \A.}
    
        \begin{tabular}{ r  r  r  c c c }          
                \hline                     
  
                  Configuration & Term & $J$ & $E$ &  $E/k_B$  \\
                  & & & ($\si{cm^{-1}}) $  & ($\si{K}$)   \\

            \hline

            3d$^6$($^5$D)4s     & a$^6$D & 9/2 & 0 & 0 \\
            &  & 7/2 & 385 & 554 \\
            &  & 5/2 & 668 & 961 \\
            &  & 3/2 & 863 & 1241 \\
            &  & 1/2 & 977 & 1406 \\

            &&&&\\

            3d$^7$      & a$^4$F & 9/2 & 1873 & 2694 \\
            &  & 7/2 & 2430 & 3497 \\
            &  & 5/2 & 2838 & 4084 \\
            &  & 3/2 & 3117 & 4486 \\

            &&&&\\
            
            3d$^6$($^5$D)4s     & a$^4$D & 7/2 & 7955 & 11448 \\
            &  & 5/2 & 8392 & 12077 \\
            &  & 3/2 & 8680 & 12492 \\
            &  & 1/2 & 8847 & 12732 \\

        \hline

        \end{tabular}

    \end{threeparttable}

    \label{list levels}
\end{table}

\section{Full studied spectra}

\newpage
\begin{figure*}[hbtp]
    \centering    
    \includegraphics[ trim = 150 20 120 30, scale = 0.35, clip]
    {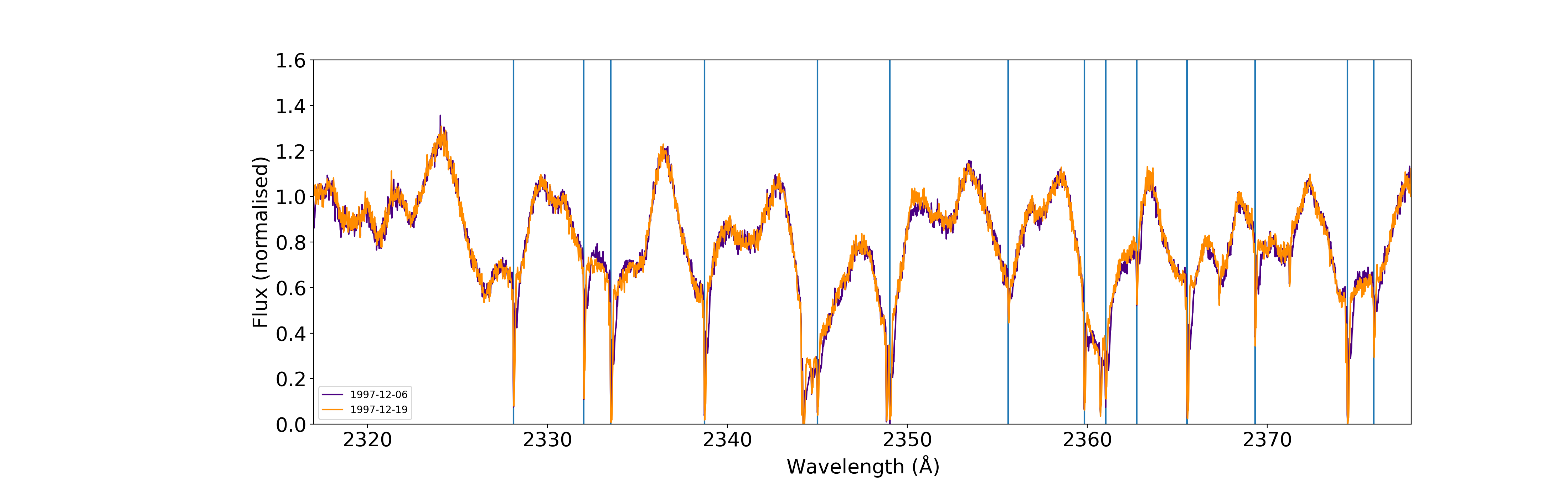}    
    \includegraphics[ trim = 150 20 115 20, scale = 0.35, clip]
    {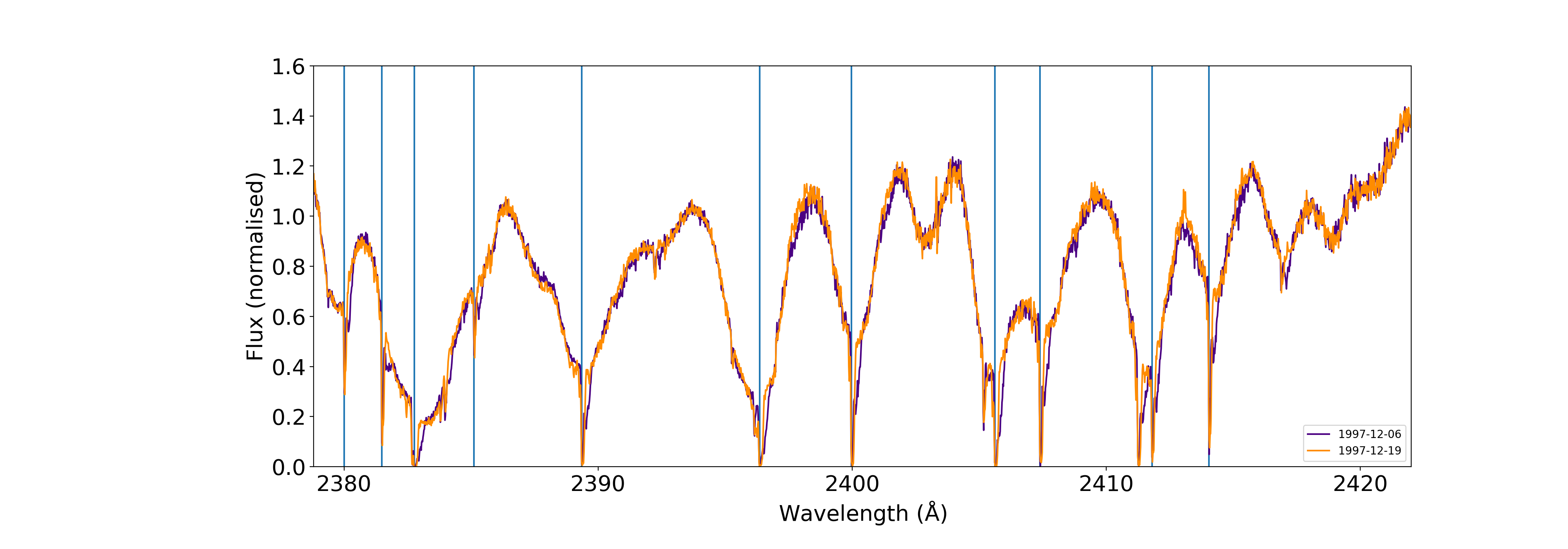}    
    \includegraphics[ trim = 250 20 145 20, scale = 0.35, clip]
    {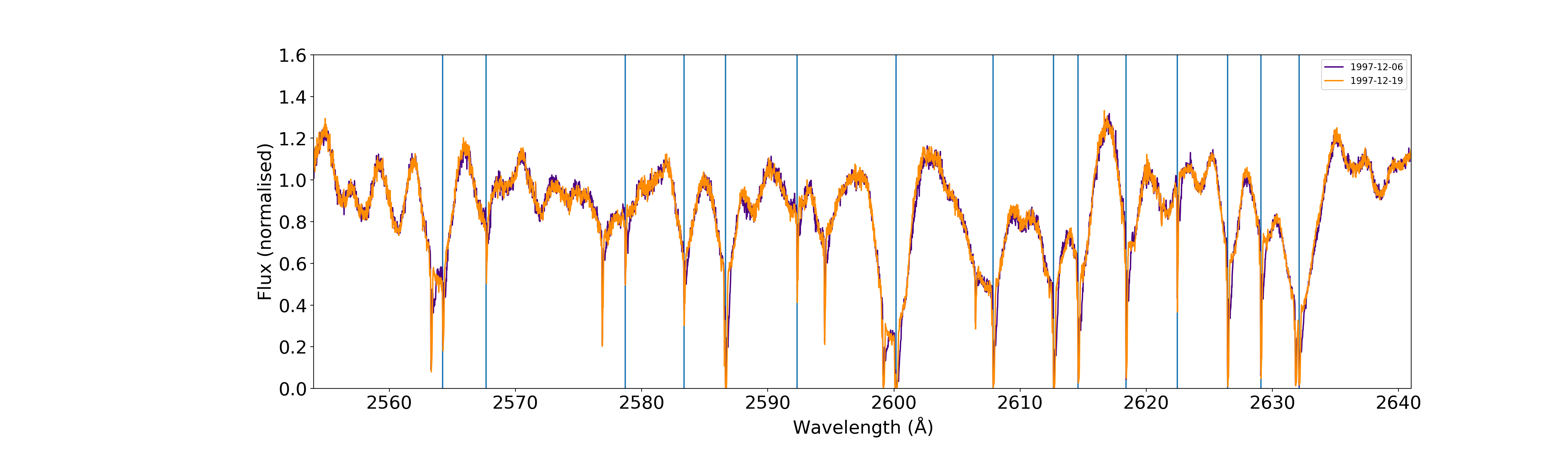}    
    \includegraphics[ trim = 250 10 145 20, scale = 0.35, clip]
    {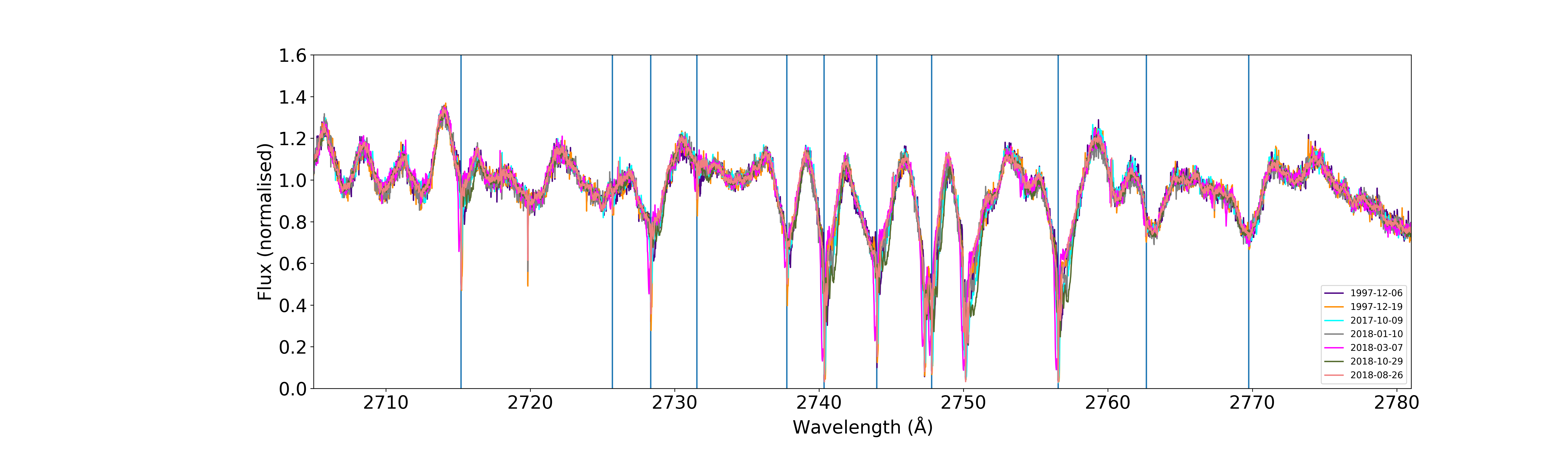}    

    \caption{Normalised spectra of \bp\ at seven different epochs, focussing on three \feii\ line series at 2400, 2600, and 1750 \A. Vertical blue lines emphasise the \feii\ transitions that were used in our analysis and are listed in Table \ref{list lines}. Other lines not used in this paper may also show cometary features, such as \feii\ multiplets (e.g. 2344, 2750 \A) or lines from other species (e.g. \mnii\ at 2576, 2594 and 2606 \A). }
    \label{Full spectrum}
\end{figure*}

\FloatBarrier

\newpage

\onecolumn

\section{Line examples}

\begin{figure*}[h!]
    \includegraphics[
    trim = 20 10 20 10, clip, scale = 0.38]
    {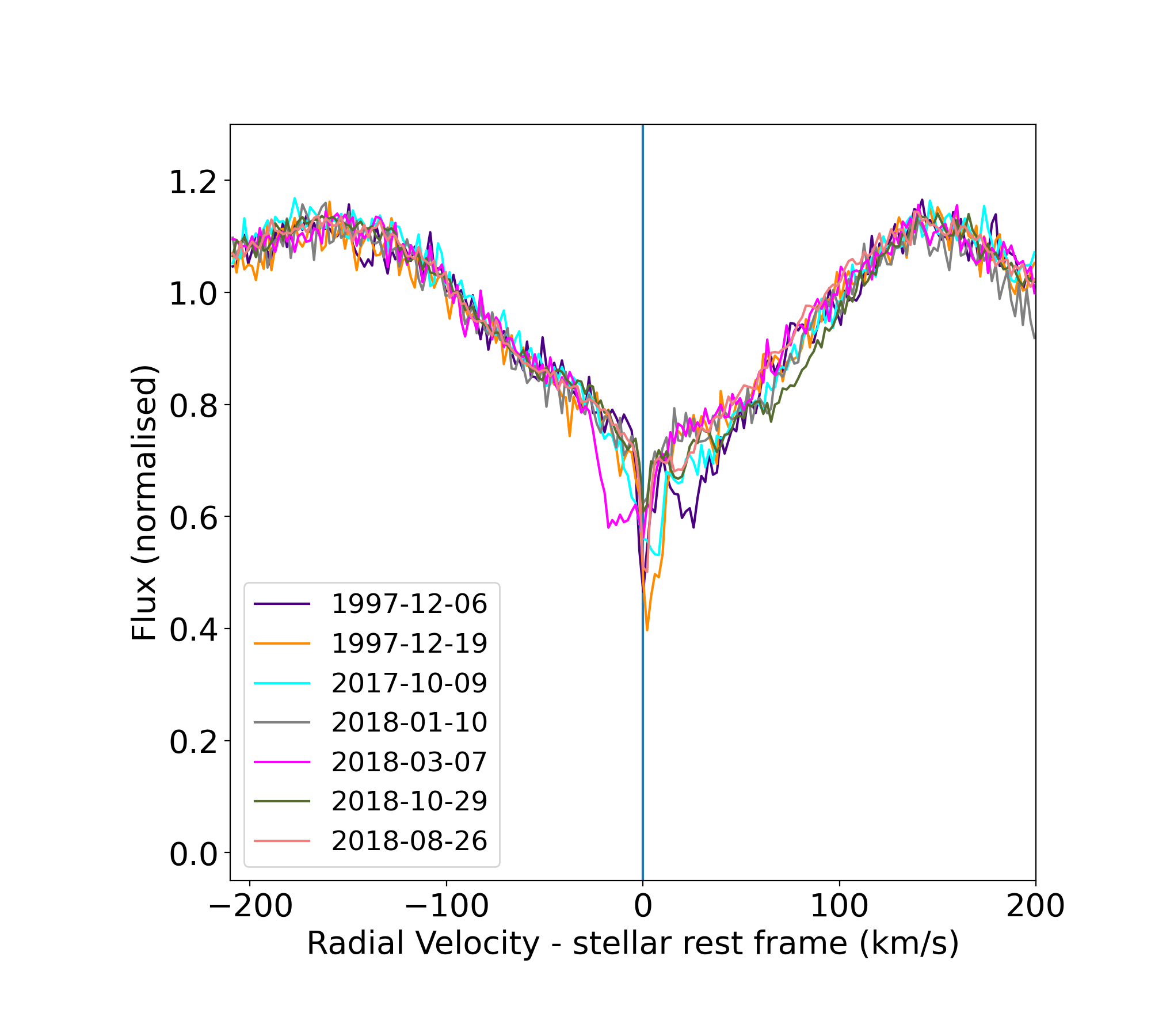}    
    \includegraphics[ 
    trim = 30 10 20 10, clip, scale = 0.38]
    {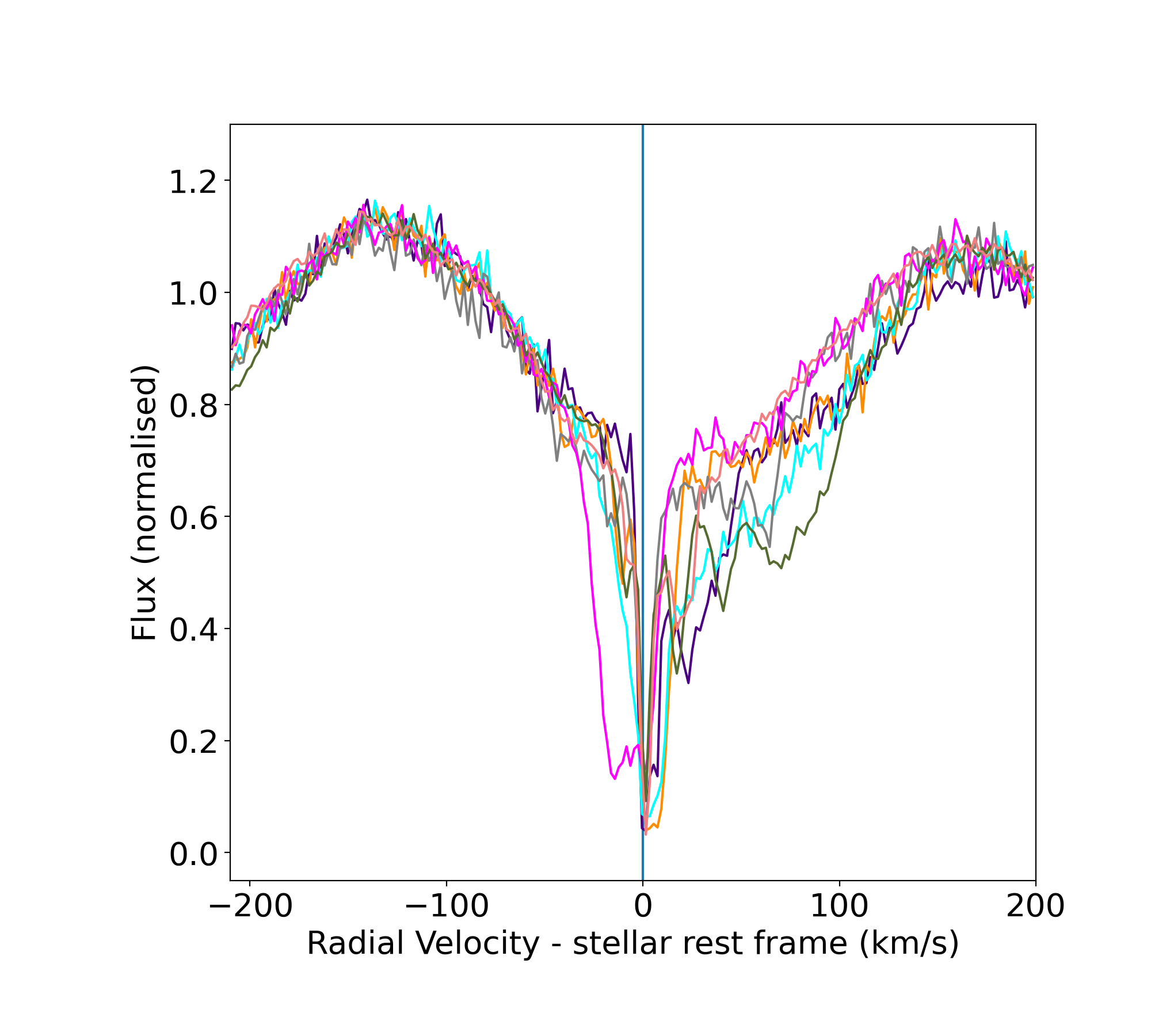}    

    \includegraphics[ 
    trim = 20 10 20 10, clip, scale = 0.38]
    {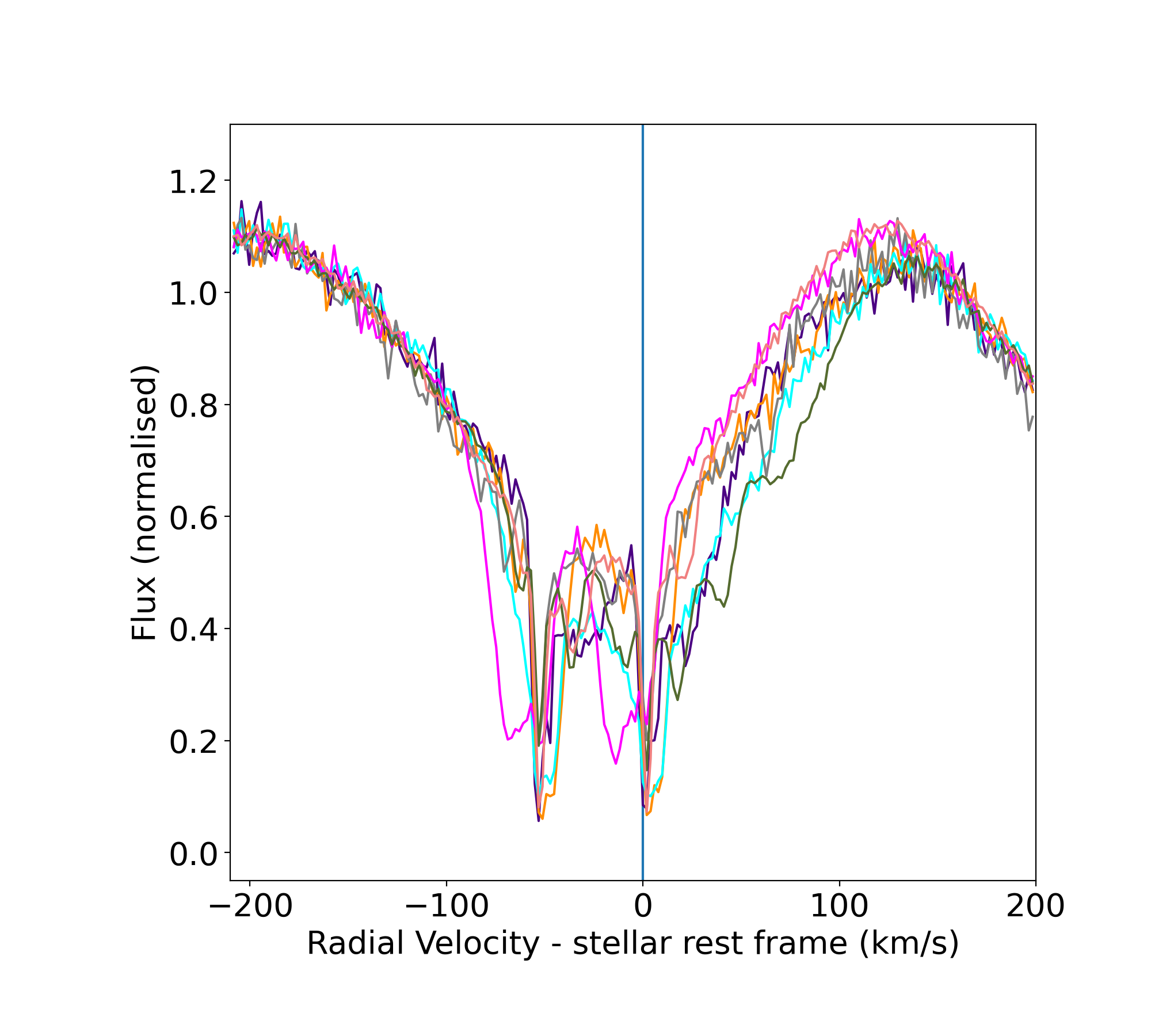}
    \includegraphics[ 
    trim = 30 10 20 10, clip, scale = 0.38]
    {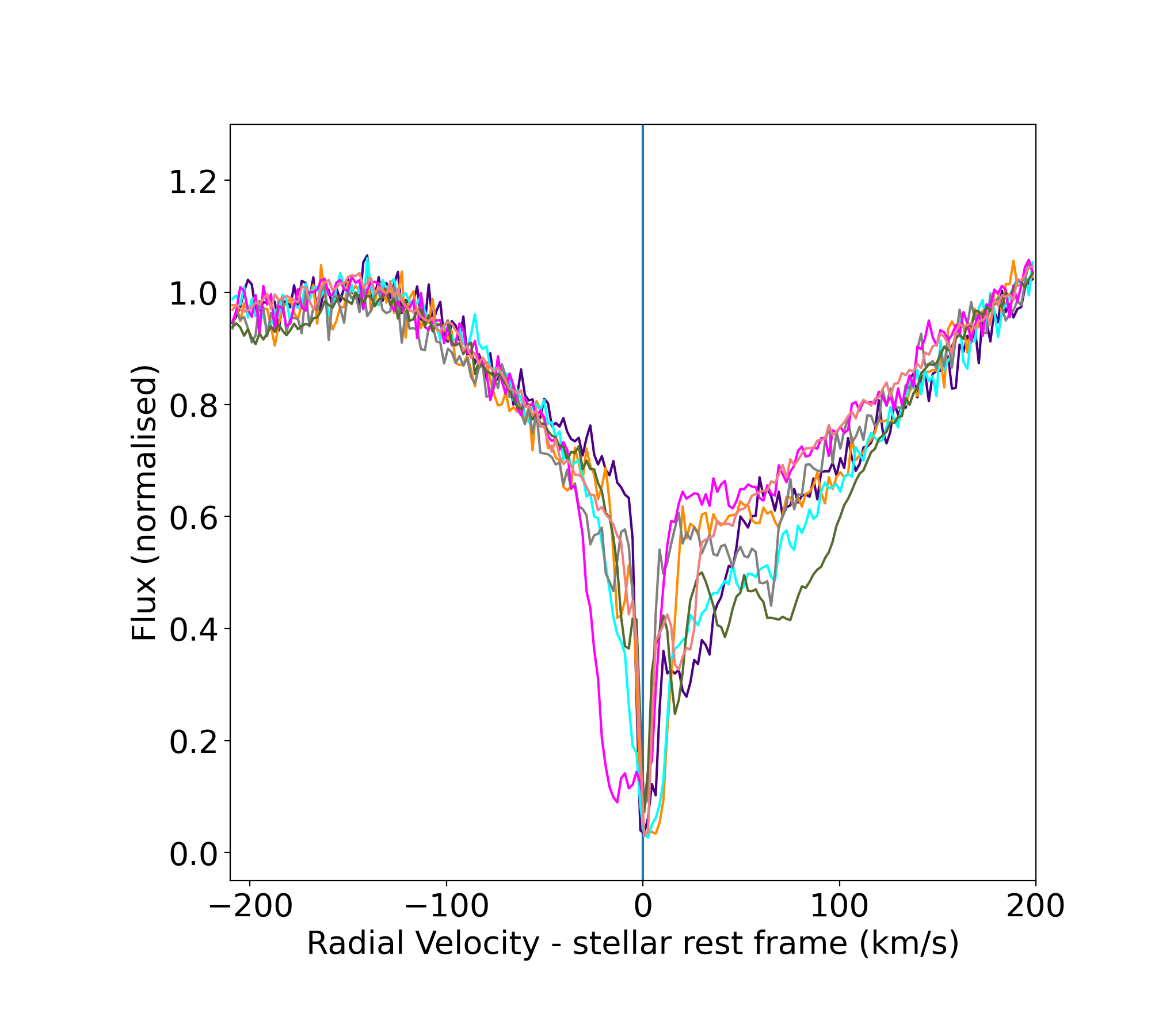}
   
    \caption{Examples of \bp\ spectra in \feii\ lines near 2750 \A, rising from a$^4$D metastable term. The lines are at 
    2737\,\AA\  (top left, $E_l = 8680 \ \si{cm^{-1}}$), 
    2740\,\AA\ (top right, $E_l = 7955 \ \si{cm^{-1}}$), 
    2747.8\,\AA\ (bottom left, $E_l = 8392 \ \si{cm^{-1}}$) 
    and 2756\,\AA\ (bottom right, $E_l = 7955 \ \si{cm^{-1}}$). The blue lines emphasise the location of the lines in the \bp\ rest frame (thus at 0 km/s).}
    \label{Fe II lines 2750}
\end{figure*}
\newpage

\onecolumn

\begin{figure*}[h]
    \includegraphics[ 
    trim = 20 10 20 10, clip, scale = 0.38]
    {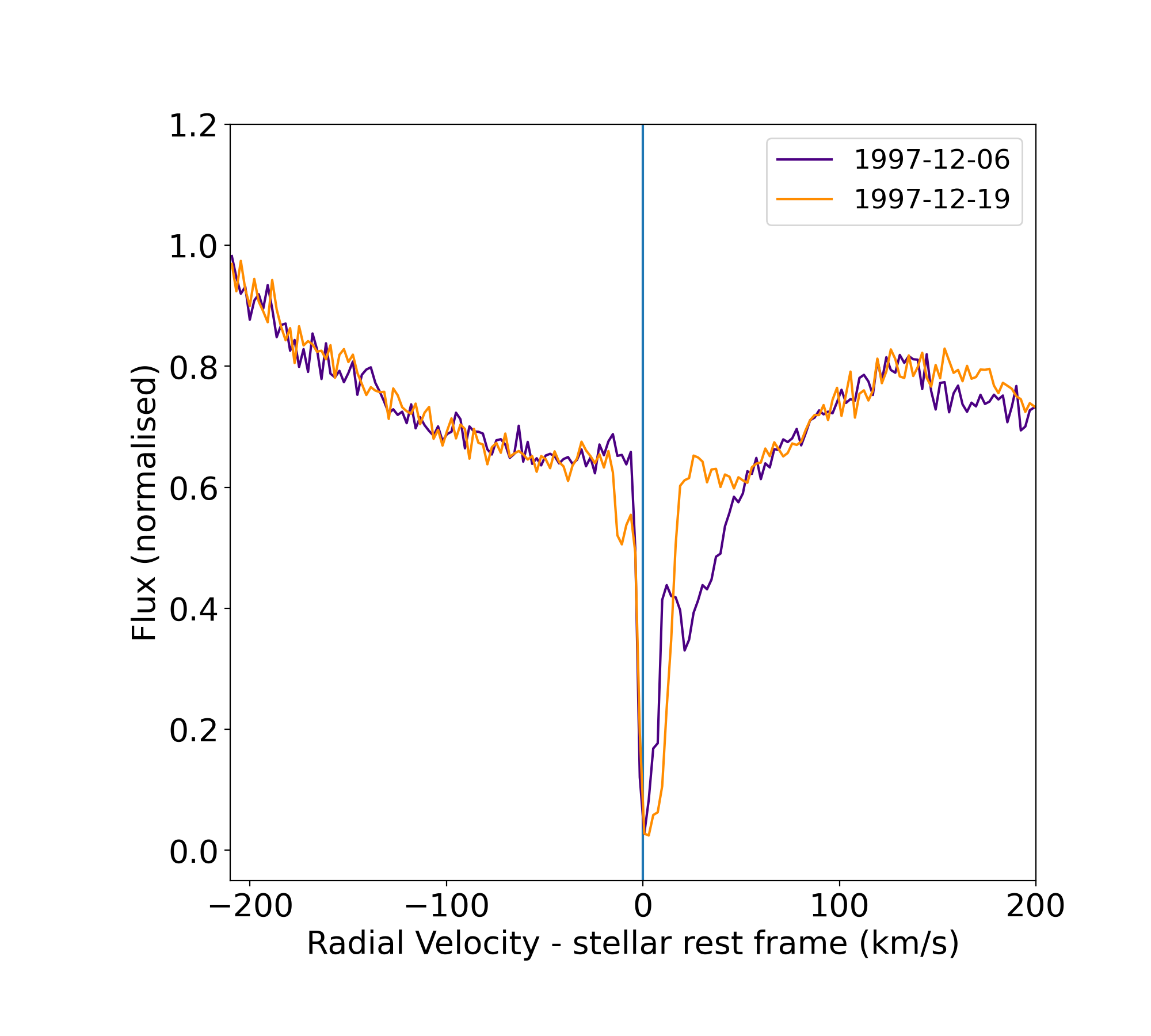}    
    \includegraphics[ 
    trim = 30 10 20 10, clip, scale = 0.38]
    {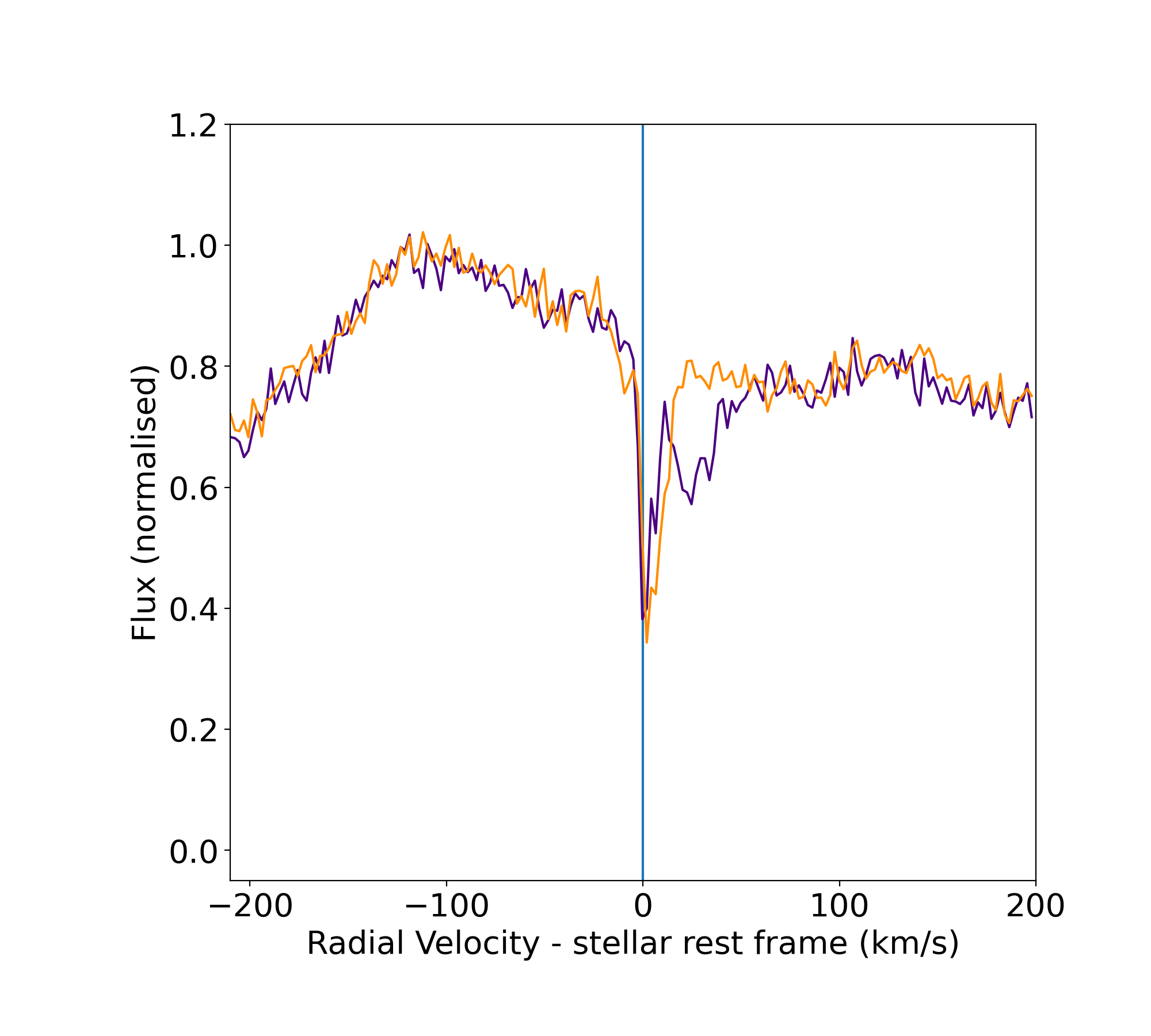}    

    \includegraphics[ 
    trim = 20 10 20 10, clip, scale = 0.38]
    {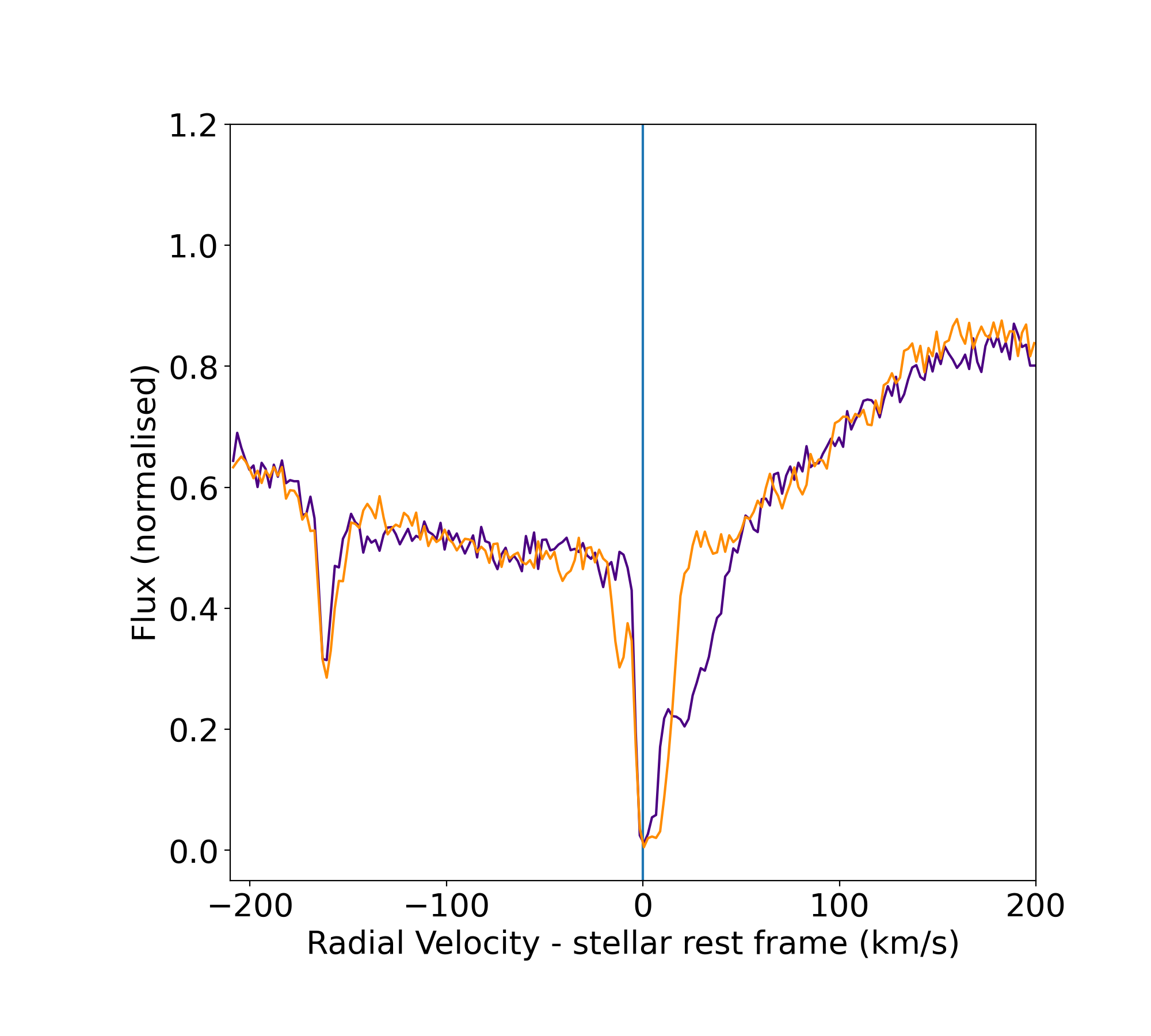}
    \includegraphics[ 
    trim = 30 10 20 10, clip, scale = 0.38]
    {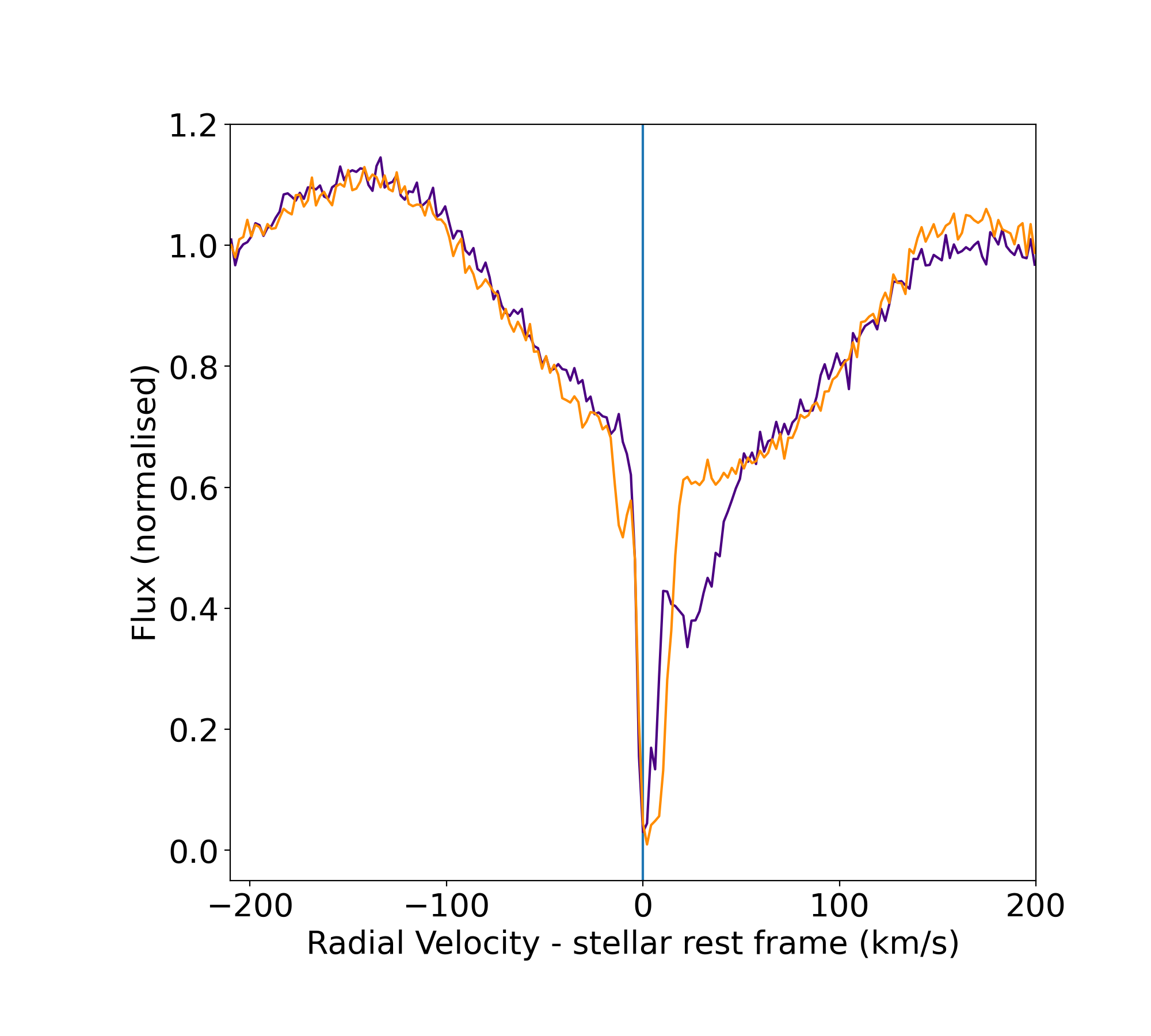}
    \caption{Examples of \feii\ lines from low energy levels in 2300-2600\,\A\ region: 
    2365\,\A\ (top-left, $E_l = 385 \si {cm^{-1}}$), 
    2369\,\A\ (top right, $E_l = 4080 \si {cm^{-1}}$), 
    2607\,\A\ (bottom left, $E_l = 668 \si {cm^{-1}}$), 
    and 2626\,\A\ (bottom right, $E_l = 385 \si {cm^{-1}}$). 
    Only two observations obtained in December 1997 are available on this wavelength range. Again, the blue lines emphasise the location of the lines in the stellar rest frame.}
    \label{Fe II lines 2600}    
\end{figure*}
\newpage

\onecolumn

\begin{figure*}[h]
    \includegraphics[ 
    trim = 20 10 20 10, clip, scale = 0.38]
    {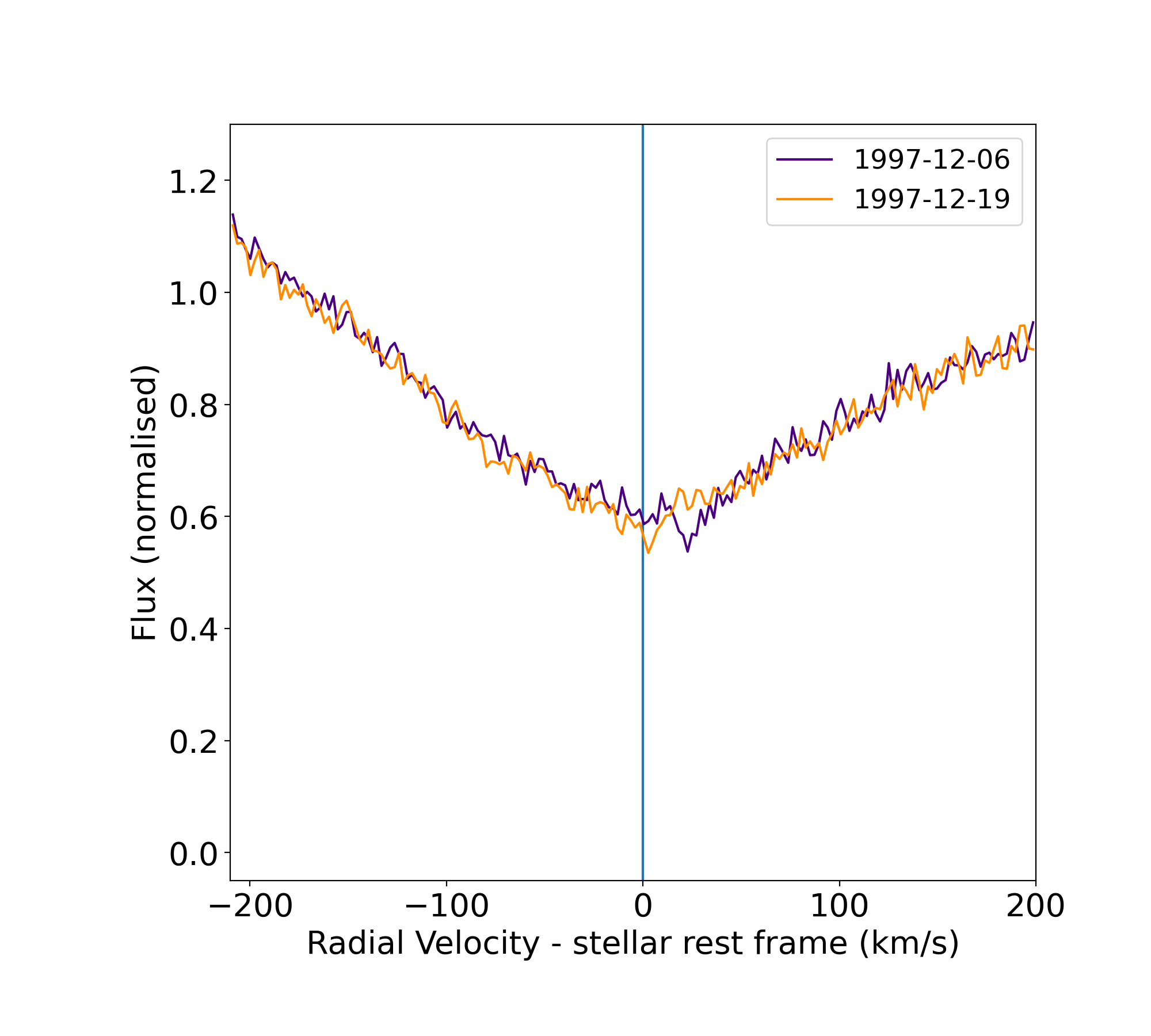}    
    \includegraphics[ 
    trim = 30 10 20 10, clip, scale = 0.38]
    {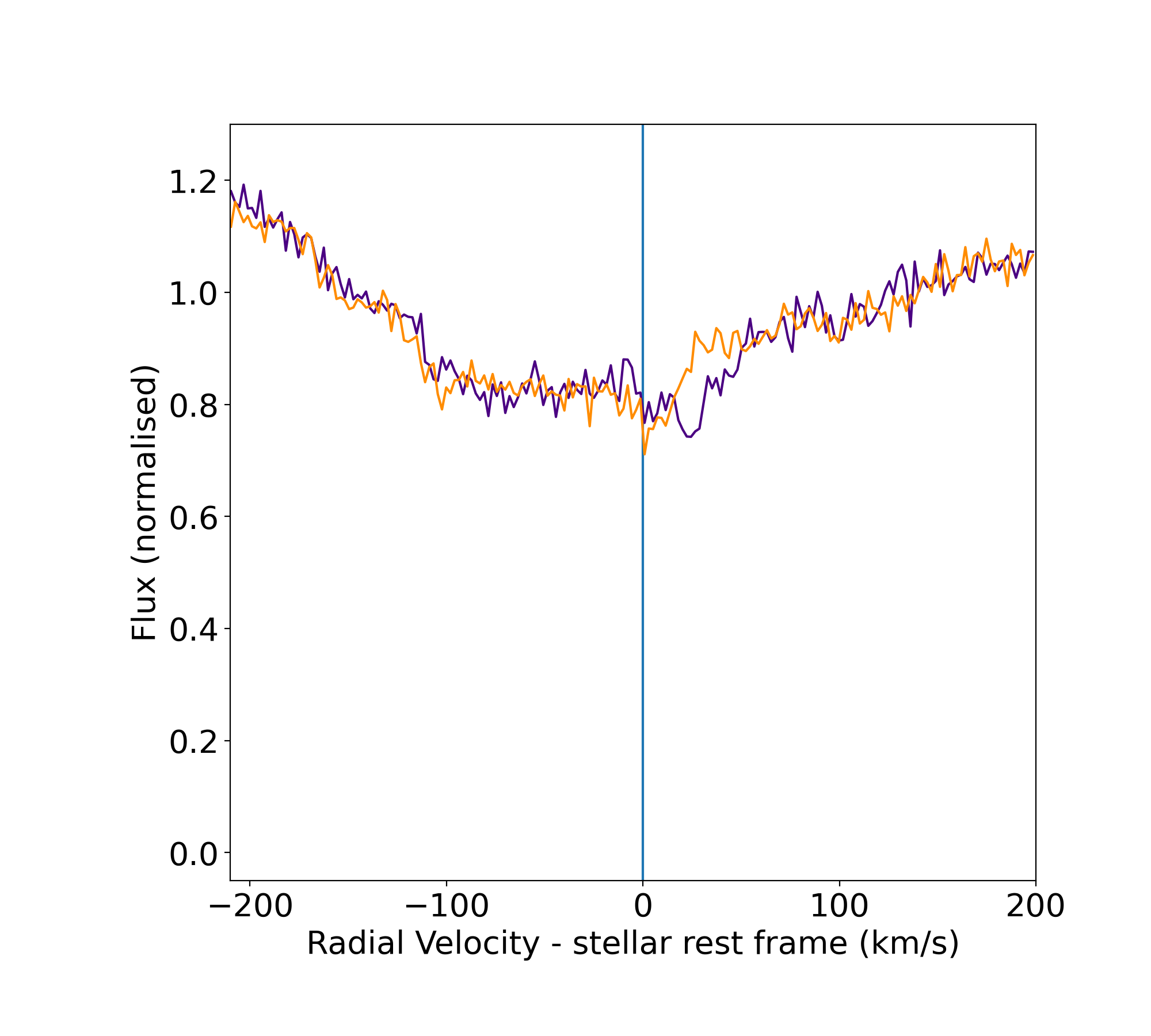}    

    \includegraphics[ 
    trim = 20 10 20 10, clip, scale = 0.38]
    {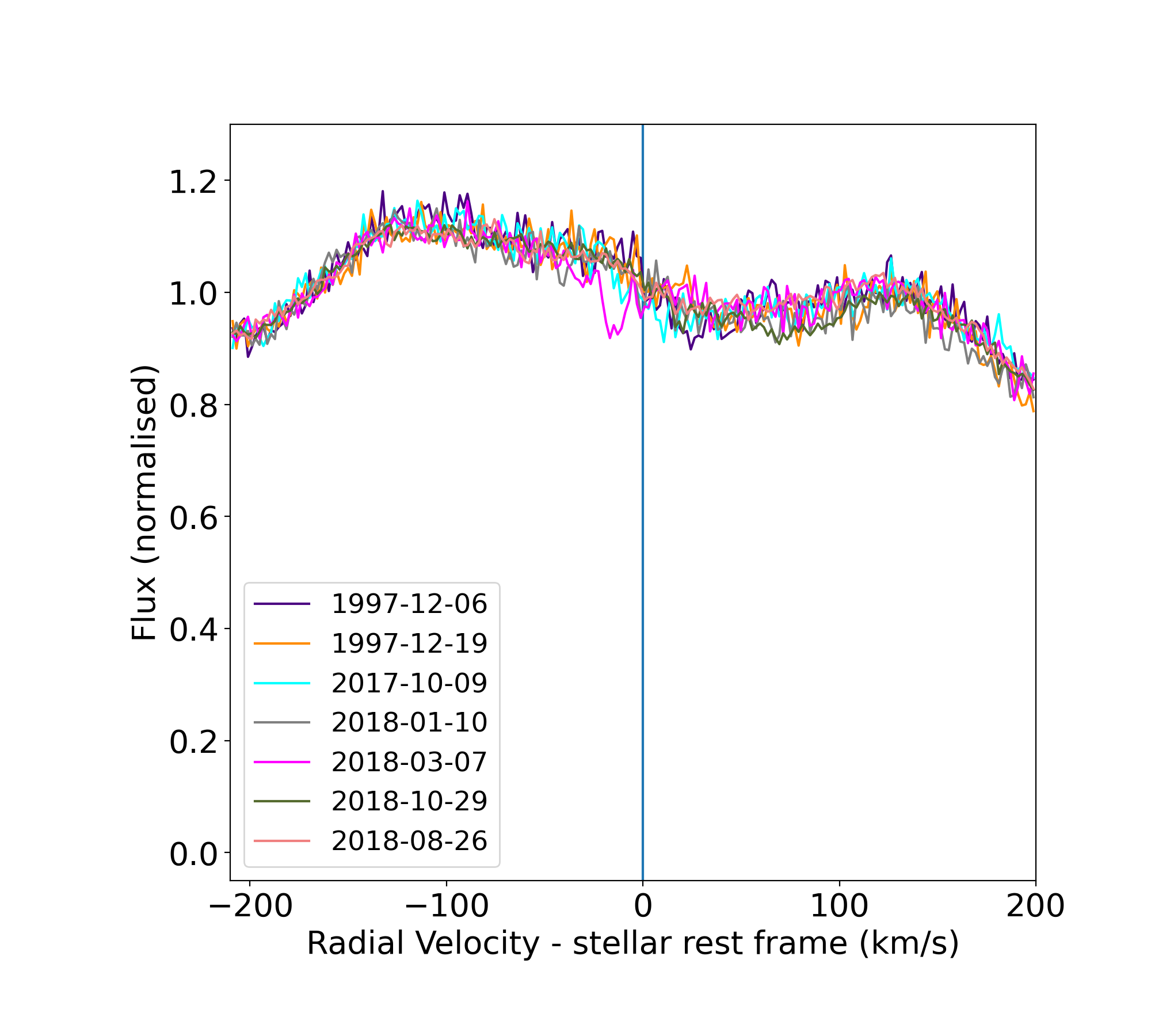}
    \includegraphics[ 
    trim = 30 10 20 10, clip, scale = 0.38]
    {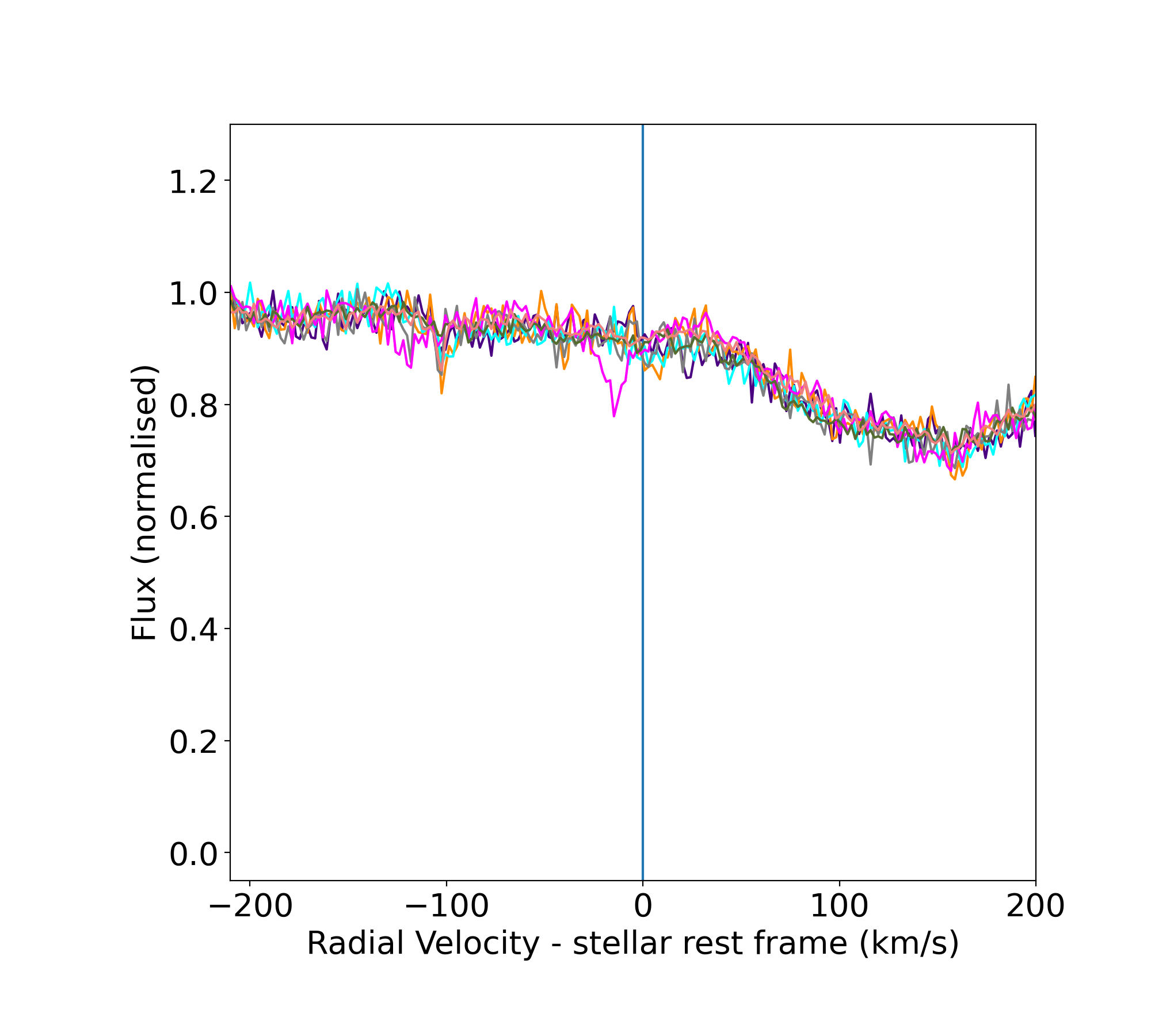}
    \caption{Examples of faint \feii\ lines rising from highly excited levels: 
    2424\,\A\ (top left, $E_l = 22637 \si {cm^{-1}}$), 
    2512\,\A\ (top right, $E_l = 21712 \si {cm^{-1}}$), 
    2754\,\A\ (bottom left, $E_l = 26352 \si {cm^{-1}}$), 
    and 2768\,\A\ (bottom right, $E_l = 26170 \si {cm^{-1}}$), with blue lines indicating the location of the lines in \bp\  rest frame.}
    \label{Fe II lines faint}
    \vspace{0.3 cm}
\end{figure*}

\end{appendix}
\end{document}